      [1999/12/01 v1.4c Il Nuovo Cimento]
\documentclass{cimento}

\usepackage{graphicx}  
\title{Simulating Galaxy Clusters}
\author{Michael L. Norman\from{ins:x}}
\instlist{\inst{ins:x} Physics Department and CASS, UC San Diego, USA}
\begin{document}

\maketitle

\section{Introduction}
Galaxy clusters are the largest gravitationally bound objects in the universe. 
They are also among the rarest objects in the universe. While these two facts
about galaxy clusters may seem disparate, they are in fact intimately related.
Our current theory of the origin and evolution of galaxy clusters places them
within the broader context of cosmological structure formation in which galaxies,
galaxy groups, galaxy clusters, and galaxy superclusters all arise from gravitational
instability amplifying perturbations in the cold dark matter density field
in an expanding universe. At early times the perturbations are linear in amplitude, 
and are extremely well described by a Gaussian random field with a known power spectrum
(the $\Lambda CDM$ power spectrum; cf. Fig. \ref{fig.ps}). 
At later times, density perturbations become nonlinear
and collapse into gravitationally bound systems. The shape of the $\Lambda CDM$ 
power spectrum is such that structure forms from the ``bottom up", with galaxies forming
first and clusters forming later. It just so happens that we live in a universe
in which cluster-scale perturbations collapsed rather recently (since $z \sim 1$),
which accounts for their rarity as well as their sometimes complex substructure.

\begin{figure}
\begin{center}
\includegraphics[width=\textwidth]{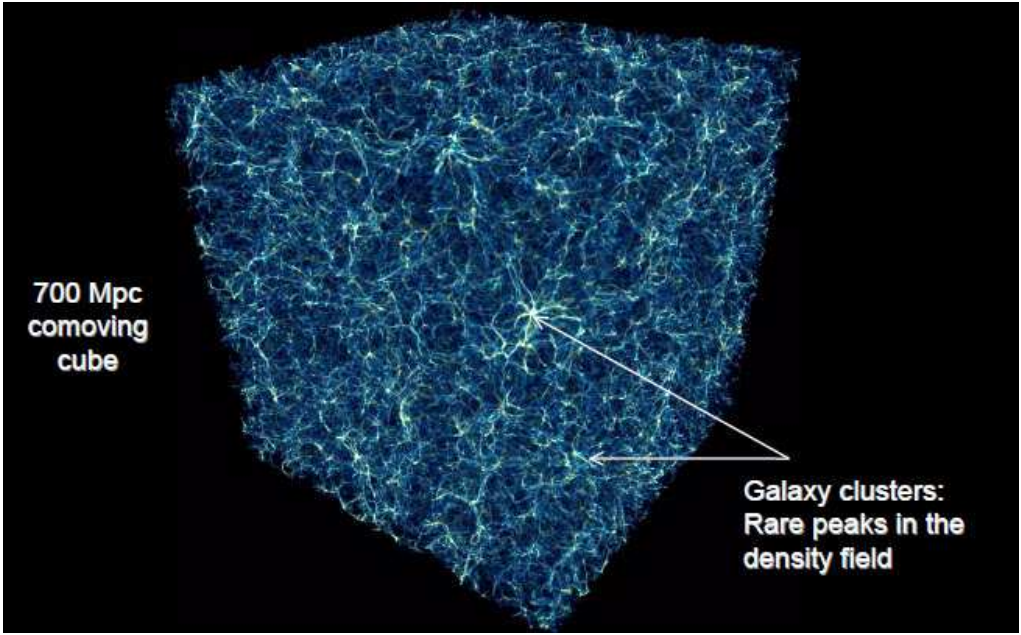}
\end{center}
\caption{AMR hydrodynamic cosmological simulation of cosmic structure in a 700 Mpc volume
of the universe. Up to seven levels of adaptive mesh refinement (AMR) resolve the distribution of baryons within and between galaxy clusters, for an effective resolution of
$65,536^3$. Volume rendering of baryon density. From \cite{Norman07}.}
\label{fig.cube}
\end{figure}

As cluster-scale perturbations collapse, they bring in all matter within a sphere 
of comoving radius of
about 15 Mpc, which includes galaxies, intergalactic gas, and anything else in
that patch of the universe. Because the escape velocity of galaxy clusters is
of order 1000 km/s, everything but relativistic particles become trapped in the
cluster potential well. For this reason it is often said that clusters represent
a fair sample of the universe. This is true from the standpoint of their matter
content. However, from the standpoint of
cosmological structure this could not be further from the truth. Galaxy clusters
form and evolve in the rarest peaks ($\sim 3\sigma$) of the density field 
(Fig. \ref{fig.cube}). Galaxy 
formation begins sooner in such regions, and the galaxies evolve due to internal and
external processes which are somewhat different from the general field (e.g., ram
pressure stripping). 

Galaxy clusters are interesting objects in their own right, and for decades have
been extensively studied in the optical, Xray, and radio wavebands \cite{Carnegie04}. 
More recently this has been extended to the microwave, infrared, and extreme UV
\cite{Taiwan03,Varenna04}, motivated in part by the fact that galaxy clusters are
excellent cosmological probes. Because of their large size and high X-ray 
luminosities, they can be seen to great distances. As discussed in this volume and
in \cite{Varenna04}, galaxy clusters can also be seen in absorption/emission against the
cosmic microwave background (CMB) via the Sunyaev-Zeldovich effect (SZE). As discussed
by Rephaeli elsewhere in these proceedings, the SZE is redshift independent, meaning
that deep microwave surveys should detect all galaxy clusters in a particular region
of the sky regardless of their redshift provided the telescope has enough angular
resolution and sensitivity. Indeed such surveys are underway now and results are expected soon.

One of the most challenging measurements in modern observational cosmology is
the dark energy equation of state which describes 
the time rate of change of the vacuum energy
density $\dot{\Lambda}$ responsible for the accelerating expansion of the universe
\cite{Perlmutter03}. Galaxy clusters were identified by the US Dark Energy Task Force
(DETF) as one of four complementary methods for doing this. However in order to measure
$\dot{\Lambda}$ we must measure the cluster mass distribution function 
versus redshift to very high accuracy. 
Accurately measuring the mass of a galaxy cluster is actually quite difficult
despite the number of ingenious techniques that have been developed. 
This motivates attempts to measure the ``mass-observable" relationships in
observed samples of clusters, and find the ones with the least scatter and least bias. 

Numerical simulations
are helpful in this regard, as one can in principle calibrate the mass-observable 
relationships by comparing simulated observations with {\em in situ} measurements. 
Where this has been done, a discouraging result is found: most of the observables of 
a given simulated cluster depend sensitively on numerical resolution and assumed
baryonic physics, including radiative cooling, star formation, galactic and AGN feedback
processes. However it has been shown that the integrated SZE signal is rather insensitive
to assumed baryonic physics \cite{Motl04,Nagai06}, perhaps relaxing requirements on modelers
and giving encouragement to SZE cluster observers that the dark energy program may be
feasible after all. The point is that the astrophysics of galaxy clusters and their
utility as cosmological probes are inextricably linked, and both are worthwhile of 
study.

I was asked by the organizers to lecture on three topics relavant to the theme of the summer school.
First I was asked to review the standard cosmological framework and basic results
from the theory of cosmological structure formation within which galaxy clusters 
can be understood. Second, I was asked to review how galaxy clusters are simulated
on a computer, and summarize the basic findings. I do this in the next two sections, which
are slightly updated and abbreviated versions of the lecture notes I published in the
2004 Varenna Summer School volume \cite{Varenna04}. Finally I was asked to give a 
lecture of my choosing, which was on recent progress in galaxy cluster modeling focusing
on the incorporation of additional baryonic physics and simulating SZE surveys . These topics
are presented in Sections 5 and 6 respectively. 

In line with the character of the summer school, I have attempted to be 
pedagogical, emphasizing the key concepts and results that a student needs 
to know if s/he wants to understand the current literature or do 
research in this area. Literature citations are kept to a minimum, except 
for textbooks, reviews, and research papers that I found to be particularly 
helpful in preparing this article. I am indebted to Dr. Rocky Kolb whose slides
much of Section 2 are based upon. 

\section {Cosmological framework and perturbation growth in the 
linear regime}

Our modern theory of the structure and evolution of the universe, along with 
the observational data which support it, is admirably presented in the
textbook by Dodelson \cite{dodelson03}. 
Remarkable observational progress has been made in the past two 
decades which has strengthened our confidence in the correctness of the 
hot, relativistic, expanding universe model (Big Bang), has measured the 
universe's present mass-energy contents and kinematics, and lent strong 
support to the notion of a very early, inflationary phase. Moreover, 
observations of high redshift supernovae unexpectedly have revealed that the 
cosmic expansion is accelerating at the present time, implying the existence 
of a pervasive, dark energy field with negative pressure \cite{Perlmutter03}. This surprising discovery has enlivened observational efforts to 
accurately measure the cosmological parameters over as large a fraction of 
the age of the universe as possible, especially over the redshift interval 0 
$<$ z $<$ 1.5 which, according to current estimates, spans the 
deceleration-acceleration transition. These efforts include large surveys of 
galaxy large scale structure, galaxy clusters, weak lensing, the Lyman alpha 
forest, and high redshift supernovae, all of which span the relevant 
redshift range. Except for the supernovae, all other techniques rely on 
measurements of cosmological structure in order to deduce cosmological 
parameters.

\subsection{Cosmological standard model}
The dynamics of the expanding universe is described by the two Friedmann 
equations derived from Einstein's theory of general relativity under the 
assumption of homogeneity and isotropy. The expansion rate at time $t$ is 
given by
\begin{equation}\label{eq1}
H^2(t)\equiv \left( {\frac{\dot {a}}{a}} \right)^2=\frac{8\pi 
G}{3}\sum\limits_i {\rho _i } -\frac{k}{a^2}+\frac{\Lambda }{3}
\end{equation}
where $H(t)$ is the Hubble parameter and $a(t)$ is the FRW scale factor at time 
$t$. The first term on the RHS is proportional to the sum over 
all energy densities in 
the universe $\rho _{i }$ including baryons, photons, neutrinos, dark 
matter and dark energy. We have explicitly pulled the dark energy term out 
of the sum and placed it in the third term assuming it is a constant (the 
cosmological constant). The second term is the curvature term, where 
$k=0,\pm 1$ for zero, positive, negative curvature, respectively. Equation (\ref{eq1}) 
can be cast in a form useful for numerical integration if we introduce 
$\Omega $ parameters:
\begin{equation}\label{eq2}
\Omega _i \equiv \frac{8\pi G}{3H^2}\rho _i ,\mbox{ }\Omega _\Lambda \equiv 
\frac{8\pi G}{3H^2}\rho _\Lambda =\frac{\Lambda }{3H^2},\mbox{ }\Omega_k 
\equiv \frac{-k}{(aH)^2}
\end{equation}
Dividing equation (\ref{eq1}) by $H^2$ we get the sum rule 1=$\Omega 
_{m}+\Omega _{k}+\Omega _{\Lambda }$, which is true at all times, 
where $\Omega _{m}$ is the sum over all $\Omega _{i}$ excluding dark 
energy. At the present time $H(t)=H_{0}, a=1$, and cosmological density 
parameters become
\begin{equation}\label{eq3}
\Omega _i (0)=\frac{8\pi G}{3H_0^2 }\rho _i (0),\mbox{ }\Omega _\Lambda 
(0)=\frac{\Lambda }{3H_0^2 },\mbox{ }\Omega _k (0)=\frac{-k}{H_0^2 }
\end{equation}
Equation (\ref{eq1}) can then be manipulated into the form
\begin{equation}\label{eq4}
\dot {a}=H_0 [\Omega _m (0)(a^{-1}-1)+\Omega _\gamma (0)(a^{-2}-1)+\Omega 
_\Lambda (0)(a^2-1)+1]^{1/2}
\end{equation}
Here we have explicitly introduced a density parameter for the background 
radiation field $\Omega _{\gamma }$ and used the fact that matter and radiation 
densities scale as a$^{-3}$ and a$^{-4}$, respectively,
and we have used the sum rule to eliminate $\Omega _{k}$. Equation (\ref{eq4}) is 
equation (\ref{eq1}) expressed in terms of the \textit{current} values of the density and Hubble 
parameters, and makes explicit the scale factor dependence of the various 
contributions to the expansion rate. In particular, it is clear that the 
expansion rate is dominated first by radiation, then by matter, and finally 
by the cosmological constant.

Current measurements of the cosmological parameters by different techniques 
\cite{Komatsu09} yield the following numbers [(0) notation 
suppressed]:
\[
\begin{array}{l}
 h\equiv H_0 /(100km/s/Mpc)\approx 0.72 \\ 
 \Omega _{total} \approx 1,\mbox{ }\Omega _\Lambda \approx 0.73\mbox{, 
}\Omega _m =\Omega _{cdm} +\Omega _b \approx 0.27,\Omega _k \approx 0 \\ 
 \Omega _b \approx 0.04,\mbox{ }\Omega _\nu \approx 0.005,\mbox{ }\Omega 
_\gamma \approx 0.00005 \\ 
 \end{array}
\]
This set of parameters is referred to as the concordance model \cite{bops99}, and describes a spatially flat, low matter density, high dark 
energy density universe in which baryons, neutrinos, and photons make a 
negligible contribution to the large scale dynamics. Most of the matter in 
the universe is cold dark matter (CDM) whose dynamics is discussed below. As 
we will also see below, baryons and photons make an important contribution 
to shaping of the matter power spectrum despite their small contribution to
the present-day energy budget. Understanding the evolution of 
baryons in nonlinear structure formation is essential to interpret X-ray and 
SZE observations of galaxy clusters. 

The second Friedmann equation relates the second time derivative of the 
scale factor to the cosmic pressure $p $ and energy density\textit{ $\rho $}
\begin{equation}\label{eq5}
\frac{\ddot {a}}{a}=-\frac{4\pi G}{3}(\rho +3p),\mbox{ }\rho =\sum\limits_i 
{\rho _i } =\rho _m +\rho _\gamma +\rho _\Lambda 
\end{equation}
$p$ and $\rho $ are related by an equation of state $p_{i}=w_{i}\rho 
_{i}$, with $w_{m}$=0, $w_{\gamma }$=1/3, and $w_{\Lambda }= -1$. We thus 
have
\begin{equation}\label{eq6}
\frac{\ddot {a}}{a}=-\frac{4\pi G}{3}(\rho _m +2\rho _\gamma -2\rho _\Lambda 
)\mbox{. }
\end{equation}
Expressed in terms of the current values for the cosmological parameters we 
have
\begin{equation}\label{eq7}
\frac{\ddot {a}}{a}=-\frac{1}{2}H_0^2 [\Omega _m (0)a^{-3}+2\Omega _\gamma 
(0)a^{-4}-2\Omega _\Lambda (0)]\mbox{. }
\end{equation}
Evaluating equation \ref{eq7} using the concordance parameters, we see the 
universe is currently accelerating $\ddot {a}\approx 0.6H_0^2 \mbox{ }$ . 
Assuming the dark energy density is a constant, the acceleration began when
\begin{equation}\label{eq8}
a\equiv \frac{1}{1+z}=\left( {\frac{\Omega _m (0)}{2\Omega _\Lambda (0)}} 
\right)^{\mbox{1/3}}\mbox{ }\approx 0.57
\end{equation}
or $z\sim 0.75$.

\subsection{The Linear power spectrum}

\begin{figure}[htbp]
\centerline{\includegraphics[width=5.18in,height=4.75in]{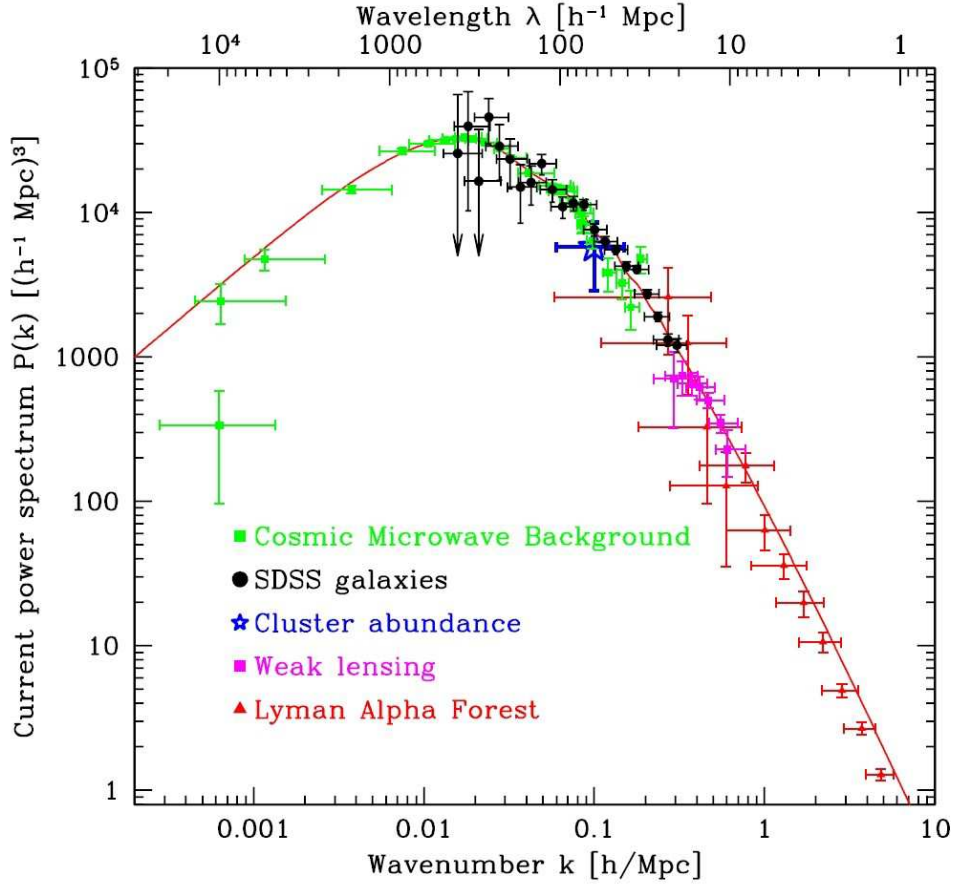}}
\caption{Linear matter power spectrum P(k) versus wavenumber extrapolated 
to z=0, from various measurements of cosmological structure. The best fit 
$\Lambda $CDM model is shown as a solid line (from \cite{tegmark03}.)}
\label{fig.ps}
\end{figure}

Cosmic structure results from the amplification of primordial density 
fluctuations by gravitational instability. The power spectrum of matter 
density fluctuations has now been measured with considerable accuracy across 
roughly four decades in scale. Figure \ref{fig.ps} shows the latest results, 
taken from reference
\cite{tegmark03}. Combined in this figure are measurements using cosmic 
microwave background (CMB) anisotropies, galaxy large scale structure, weak 
lensing of galaxy shapes, and the Lyman alpha forest, in order of decreasing 
comoving wavelength. In addition, there is a single data point for galaxy 
clusters, whose current space density measures the amplitude of the power 
spectrum on 8 h$^{-1}$ Mpc scales \cite{wef93}. 
Superimposed on the data is the predicted $\Lambda $CDM linear power 
spectrum at z=0 for the concordance model parameters. As one can see, the 
fit is quite good. In actuality, the concordance model parameters are 
determined by fitting the data. A rather complex statistical machinery 
underlies the determination of cosmological parameters, and is discussed in 
Dodelson (2003, Ch. 11). The fact that modern CMB and LSS data agree over a 
substantial region of overlap gives us confidence in the correctness of the 
concordance model. In this section, we define the power spectrum 
mathematically, and review the basic physics which determines its shape. 
Readers wishing a more in depth treatment are referred to references 
\cite{dodelson03,kolbturner90}.

At any epoch $t$ (or $a$ or $z)$ express the matter density in the universe in terms 
of a mean density and a local fluctuation:
\begin{equation}\label{eq9}
\rho (\vec {x})=\bar {\rho }(1+\delta (\vec {x}))
\end{equation}
where $\delta (\vec {x})$is the density contrast. Expand $\delta (\vec 
{x})$ in Fourier modes:
\begin{equation}\label{eq10}
\delta (\vec {x})\equiv \frac{\rho (\vec {x})-\bar {\rho }}{\bar {\rho 
}}=\int {\delta (\vec {k})\exp (-i\vec {k}\cdot \vec {x})d^3} k.
\end{equation}
The autocorrelation function of $\delta (\vec {x})$ defines the power 
spectrum through the relations
\begin{equation}\label{eq11}
\left\langle {\delta (\vec {x})\delta (\vec {x})} \right\rangle 
=\int\limits_0^\infty {\frac{dk}{k}} \frac{k^3\left| {\delta ^2(\vec {k})} 
\right|}{2\pi ^2}=\int\limits_0^\infty {\frac{dk}{k}} \frac{k^3P(k)}{2\pi 
^2}=\int\limits_0^\infty {\frac{dk}{k}} \Delta ^2(k)
\end{equation}
where we have the definitions
\begin{equation}\label{eq12}
P(k)\equiv \left| {\delta ^2(\vec {k})} \right|,\mbox{ and }\Delta 
^2(k)\equiv \frac{k^3P(k)}{2\pi ^2}.
\end{equation}
The quantity $\Delta ^2(k)$ is called the dimensionless power spectrum and is 
an important function in the theory of structure formation. $\Delta 
^2(k)$ measures the contribution of perturbations per unit logarithmic 
interval at wavenumber $k$ to the variance in the matter density fluctuations. 
The $\Lambda $CDM power spectrum asymptotes to $P(k)\sim k^{1}$ for small 
$k$, and $P(k)\sim k^{-3}$ for large $k$, with a peak a $k^{\star}\sim 2\times 10^{-2}$ h 
Mpc$^{-1}$ corresponding to $\lambda^{\star}\sim $350 h$^{-1}$ Mpc. 
$\Delta ^2(k)$ is thus asymptotically flat at high $k$, but drops off as 
$k^{4}$ at small $k$. We therefore see that most of the variance
in the cosmic density field in the universe at the present epoch is on 
scales $\lambda <  \lambda^{\star}.$ 

\begin{figure}[htbp]
\centerline{\includegraphics[width=4in,height=3in]{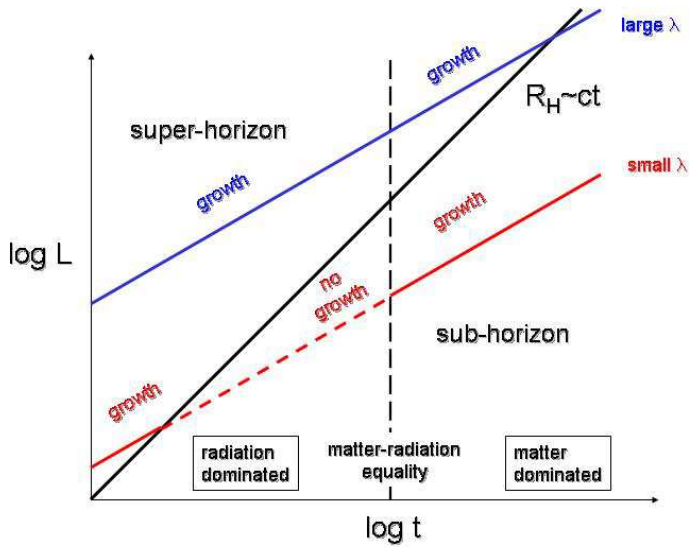}}
\caption{The tale of two fluctuations. A fluctuation which is superhorizon 
scale at matter-radiation equality grows always, while a fluctuation which 
enters the horizon during the radiation dominated era stops growing in 
amplitude until the matter dominated era begins.}
\label{fig2}
\end{figure}

What is the origin of the power spectrum shape? Here we review the basic ideas. 
Within the inflationary paradigm, it is believed that quantum mechanical 
(QM) fluctuations in the very early universe were stretched to macroscopic 
scales by the large expansion factor the universe underwent during 
inflation. Since QM fluctuations are random, the primordial density 
perturbations should be well described as a Gaussian random field. 
Measurements of the Gaussianity of the CMB anisotropies \cite{komatsu03} have confirmed 
this. The primordial power spectrum is parameterized as a power law $P_p 
(k)\propto k^n$, with $n=1$ corresponding to scale-invariant spectrum 
proposed by Harrison and Zeldovich on the grounds that any other value would 
imply a preferred mass scale for fluctuations entering the Hubble horizon. 
Large angular scale CMB anisotropies measure the primordial power spectrum 
directly since they are superhorizon scale. Observations with the WMAP 
satellite yield a value very close to $n=1$ \cite{Komatsu09}. 

To understand the origin of the spectrum, we need to understand how the 
amplitude of a 
fluctuation of fixed comoving wavelength $\lambda$ grows with 
time. Regardless of its wavelength, the fluctuation will pass through 
the Hubble horizon as illustrated in Fig. \ref{fig2}. This is because the Hubble 
radius grows linearly with time, while the proper wavelength a$\lambda $ 
grows more slowly with time. It is easy to show from Eq. \ref{eq1} that in 
the radiation-dominated era, $a\sim t^{1/2}$, and in the matter-dominated era 
(prior to the onset of cosmic acceleration) $a\sim t^{2/3}$. Thus, inevitably, 
a fluctuation will transition from superhorizon to subhorizon scale. We are 
interested in how the amplitude of the fluctuation evolves during these two 
phases. Here we merely state the results of perturbation theory (e.g., 
Dodelson 2003, Ch. 7).

\begin{figure}[htbp]
\begin{center}
\begin{tabular}{c}
\includegraphics[width=4in,height=2in]{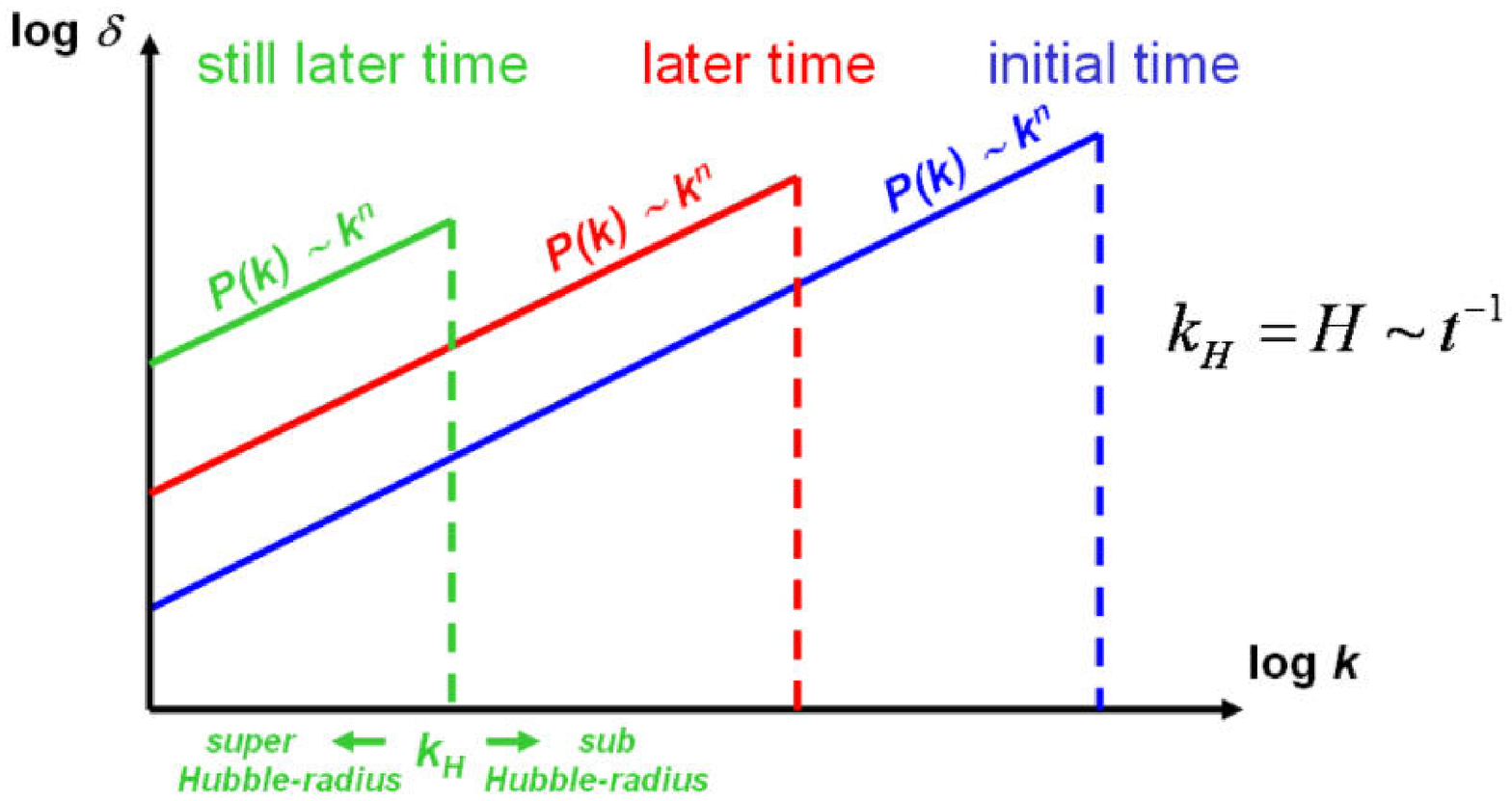} \\
\includegraphics[width=4in,height=2in]{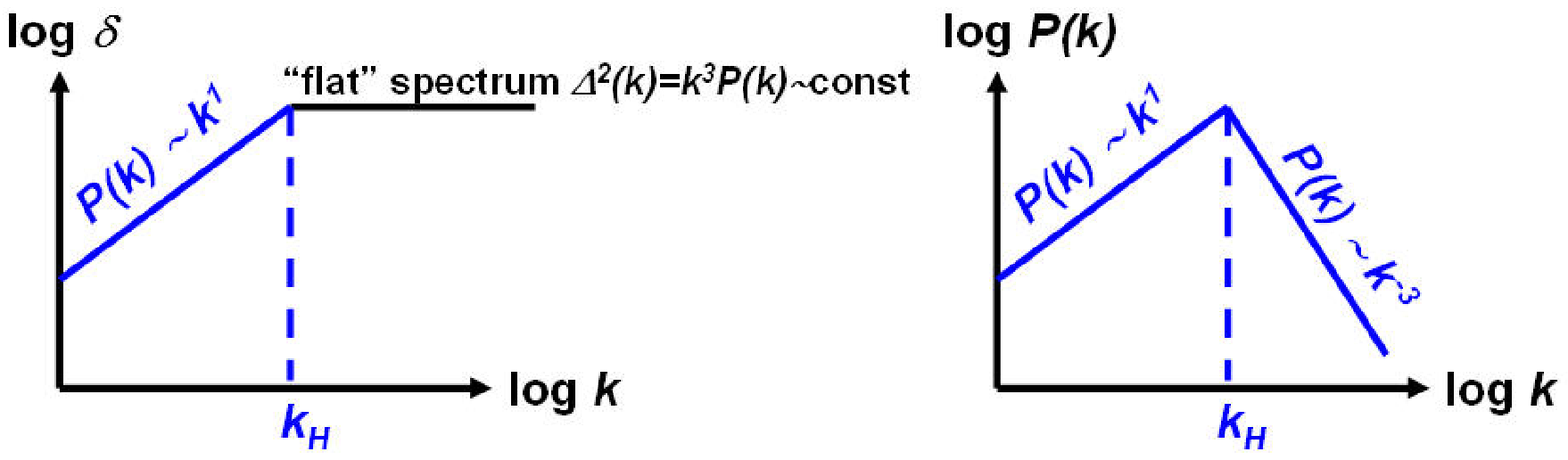}
\end{tabular}
\end{center}
\caption{ a) Evolution of the primordial power spectrum on superhorizon 
scales during the radiaton dominated era. b) Scale-free spectrum produces a 
constant contribution to the density variance per logarithmic wavenumber 
interval entering the Hubble horizon (no preferred scale) c) resulting 
matter power spectrum, super- and sub-horizon. Figures courtesy Rocky Kolb.}
\label{fig3}
\end{figure}

\subsection{Growth of fluctuations in the linear regime }

To calculate the growth of superhorizon scale fluctuations requires general relativistic 
perturbation theory, while subhorizon scale perturbations can be analyzed 
using a Newtonian Jeans analysis. We are interested in scalar 
density perturbations, because these couple to the stress tensor of the 
matter-radiation field. Vector perturbations (e.g., fluid turbulence)
are not sourced by the 
stress-tensor, and decay rapidly due to cosmic expansion. Tensor 
perturbations are gravity waves, and also do not couple to the 
stress-tensor. A detailed analysis for the scalar perturbations yields the 
following results. In the \underline {radiation dominated era}, 
\[
\begin{array}{l}
 \delta _+ (t)=\delta _+ (t_i )(t/t_i )\mbox{ superhorizon scales} \\ 
 \delta _+ (t)=constant                \mbox{ ~~~subhorizon scales} \\ 
 \end{array}
\]
while in the \underline {matter dominated era}, 
\[
\begin{array}{l}
 \delta _+ (t)=\delta _+ (t_i )(t/t_i )^{2/3}\mbox{ superhorizon scales} \\ 
 \delta _+ (t)=\delta _+ (t_i )(t/t_i )^{2/3}\mbox{ subhorizon scales} \\ 
 \end{array}
\]
This is summarized in Fig. \ref{fig2}, where we consider two fluctuations of 
different comoving wavelengths, which we will call large and small. The 
large wavelength perturbation remains superhorizon through matter-radiation 
equality (MRE), and enters the horizon in the matter dominated era. Its amplitude 
will grow as $t$ in the radiation dominated era, and as $t^{2/3}$ in the matter 
dominated era. It will continue to grow as $t^{2/3}$ after it becomes subhorizon 
scale. The small wavelength perturbation becomes subhorizon before 
MRE. Its amplitude will grow as $t$ while it is 
superhorizon scale, remain constant while it is subhorizon during the 
radiation dominated era, and then grow as $t^{2/3}$ during the matter-dominated 
era. 

Armed with these results, we can understand what is meant by a scale-free 
primordial power spectrum (the Harrison-Zeldovich power spectrum.) We are 
concerned with perturbation growth in the very early universe during the 
radiation dominated era. Superhorizon scale perturbation amplitudes grow as 
$t$, and then cease to grow after they have passed through the Hubble horizon. 
We can define a Hubble wave number $k_H \equiv 2\pi /R_H \propto 
t^{-1}.$ Fig. 4a shows the primordial power spectrum at three instants in time 
for k$<$k$_{H}$. We see that the fluctuation amplitude at k=k$_{H}$(t) 
depends on primordial power spectrum slope n. The scale-free spectrum is the 
value of n such that $\Delta ^2(k_{H}(t))$=constant for k$>$k$_{H}$. A simple 
analysis shows that this implies n=1. Since $\Delta ^2(k)\propto k^3P(k)$, we 
then have
\[
\begin{array}{l}
 P(k)\propto k^1,\mbox{ }k\le k_H \\ 
 P(k)\propto k^{-3},\mbox{ }k>k_H \\ 
 \end{array}
\]

In actuality, the power spectrum has a smooth maximum, rather than a peak as 
shown in Fig. 4c. This smoothing is caused by the different rates of growth 
before and after matter-radiation equality. 
The transition from radiation to matter-dominated is not 
instantaneous. Rather, the expansion rate of the universe changes smoothly 
through equality, as given by Eq. 1, and consequently so do the temporal 
growth rates. The position of the peak of the power spectrum is sensitive to 
the time when the universe reached matter-radiation equality, and hence is a 
probe of $\Omega _\gamma /\Omega _m $.

\begin{figure}[htbp]
\includegraphics[width=2.5in,height=1.7in]{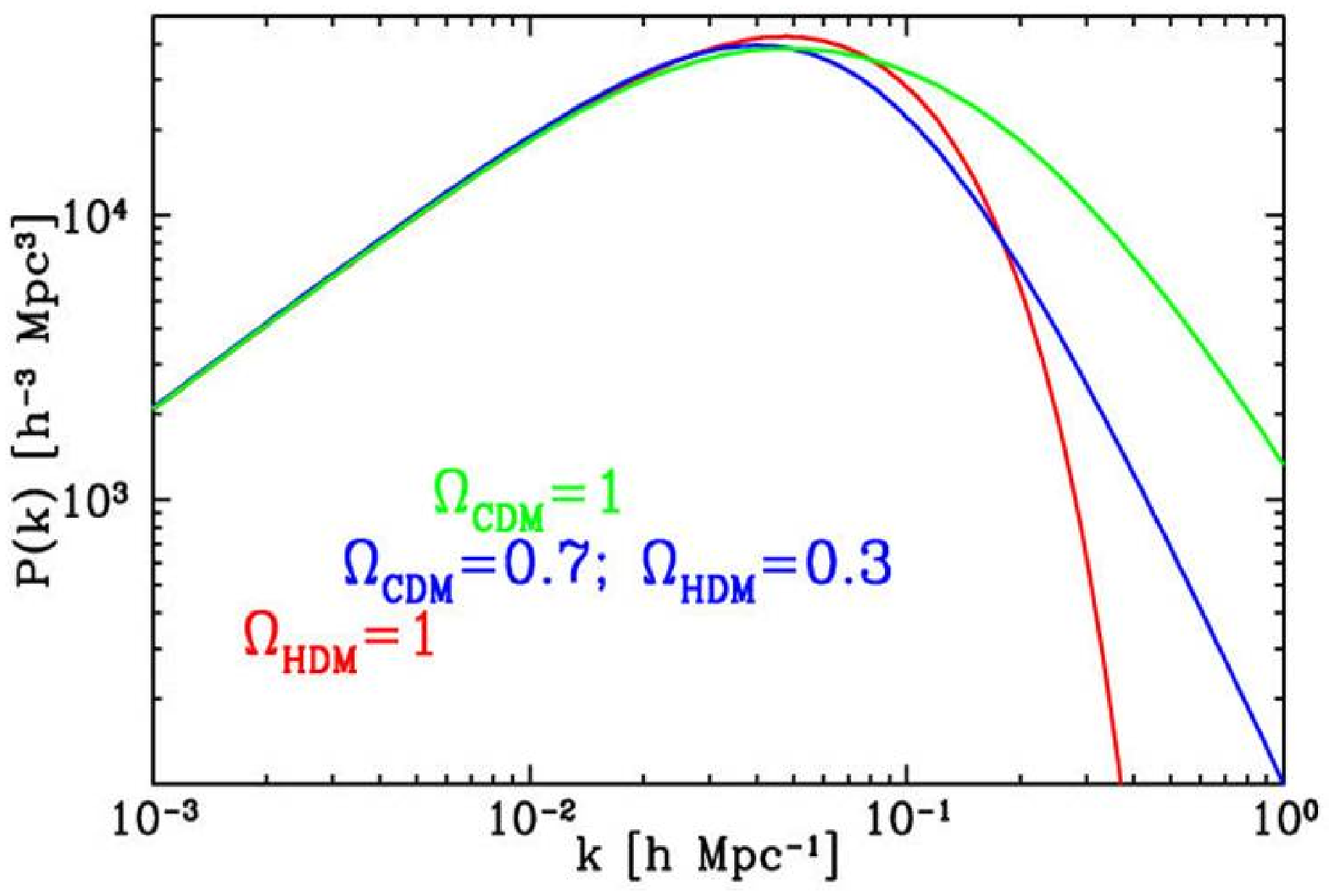}
\includegraphics[width=2.5in,height=1.7in]{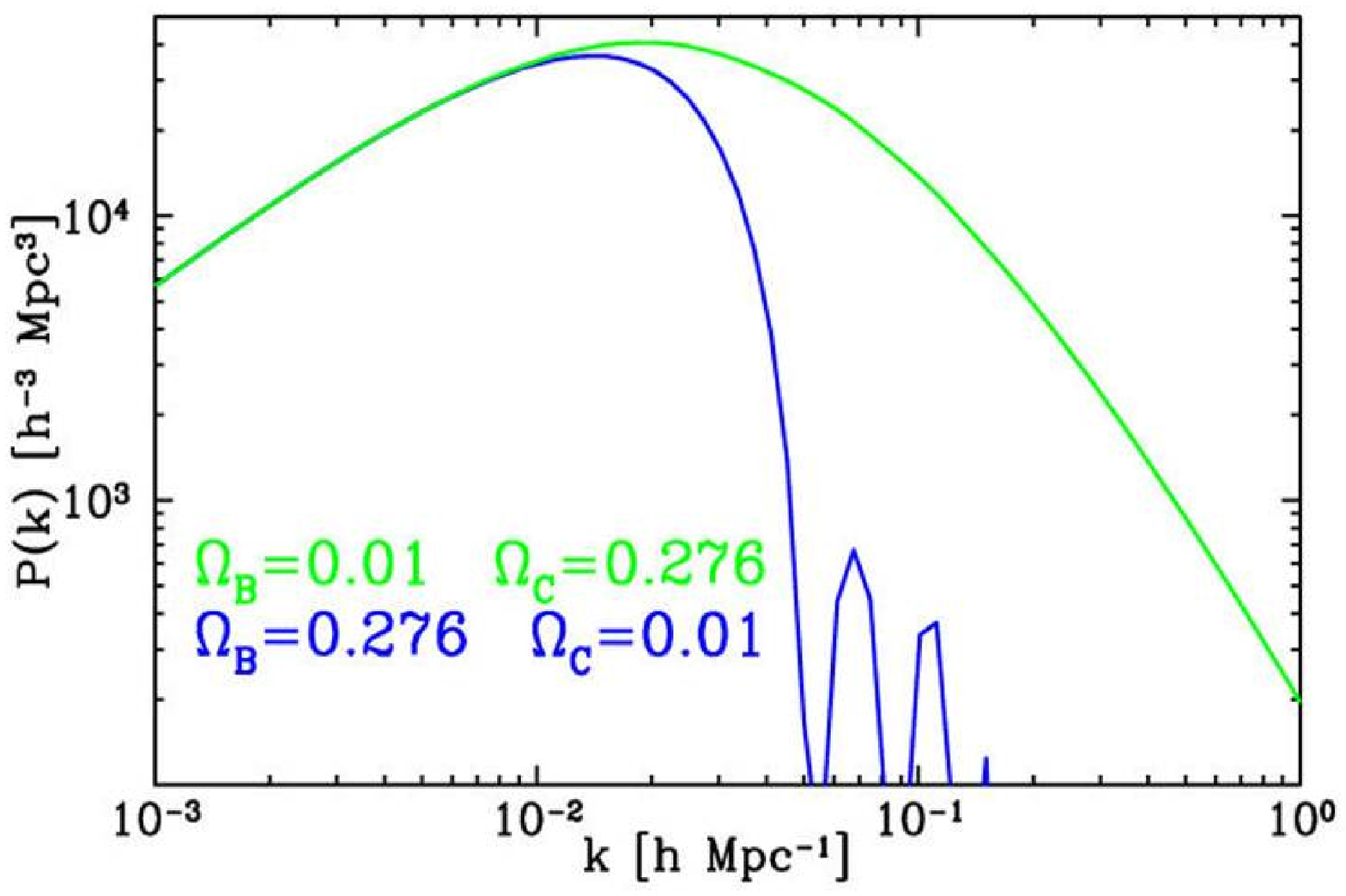}
\caption{Effect of dissipative processes on the evolved power spectrum. Left: Effect of collisionless damping (free streaming) in the dark matter. Right: Effect of collisional damping (Silk damping) in the matter-radiation fluid. Figures courtesy Rocky Kolb.}
\label{fig4}
\end{figure}

Once a fluctuation becomes sub-horizon, dissipative processes modify the 
shape of the power spectrum in a scale-dependent way. Collisionless matter 
will freely stream out of overdense regions and smooth out the 
inhomogeneities. The faster the particle, the larger its free streaming 
length. Particles which are relativistic at MRE, such as light neutrinos, 
are called hot dark matter (HDM). They have a large free-streaming length, 
and consequently damp the power spectrum over a large range of k. Weakly 
Interacting Massive Particles (WIMPs) which are nonrelativistic at MRE, are 
called cold dark matter (CDM), and modify the power spectrum very little
(Fig. \ref{fig4}). 
Baryons are tightly coupled to the radiation field by electron 
scattering prior to recombination. During rcombination, the photon mean-free 
path becomes large. As photons stream out of dense regions, they drag 
baryons along, erasing density fluctuations on small scales. This process is 
called Silk damping, and results in damped oscillations of the baryon-photon 
fluid once they become subhorizon scale. The magnitude of this effect is 
sensitive to the ratio of baryons to collisionless matter, as shown in Fig. 
\ref{fig4}. 

\section {Analytic models for nonlinear growth, virial scaling \\ 
relations, and halo statistics}

Here we introduce a few concepts and analytic results from the theory of 
structure formation which underly the use of galaxy clusters as cosmological 
probes. These provide us with the vocabulary which pervades the literature on 
analytic and numerical models of galaxy cluster evolution. Material in this 
section has been derived from three primary sources: Padmanabhan (1993) 
\cite{pad93} for 
the spherical top-hat model for nonlinear collapse, Dodelson (2003) 
\cite{dodelson03} for 
Press-Schechter theory, and Bryan {\&} Norman (1998) \cite{BN98}
for virial scaling relations.

\subsection{Nonlinearity defined}

In the linear regime, both super- and sub-horizon scale perturbations grow 
as $t^{2/3}$ in the matter-dominated era. This means that after recombination, 
the linear power spectrum retains its shape while its amplitude grows as 
$t^{4/3}$ before the onset of cosmic acceleration (Fig. 6a). When $\Delta ^2(k)$ for a 
given k approaches unity linear theory no longer applies, and some other 
method must be used to determine the fluctuation's growth. In general, 
numerical simulations are required to model the nonlinear phase of growth 
because in the nonlinear regime, the modes do not grow independently. 
Mode-mode coupling modifies both the shape and amplitude of the power 
spectrum over the range of wavenumbers that have gone nonlinear. 

At any given time, there is a critical wavenumber which we shall call the 
nonlinear wavenumber k$_{nl}$ which determines which portion of the spectrum 
has evolved into the nonlinear regime. Modes with k$<$k$_{nl}$ are said to 
be linear, while those for which k$>$ k$_{nl}$ are nonlinear (Fig. 6b). 
Conventionally, one defines the nonlinear wavenumber such that $\Delta 
(k_{nl} ,z)=1.$ From this one can derive a nonlinear mass scale $M_{nl} 
(z)=\frac{4\pi }{3}\bar {\rho }(z)\left( {\frac{2\pi }{k_{nl} }} \right)^3$. 
A more useful and rigorous definition of the nonlinear mass scale comes from 
evaluating the amplitude of mass fluctuations within spheres or radius R at 
epoch z. The enclosed mass is $M=\frac{4\pi }{3}\bar {\rho }(z)R^3.$ The mean 
square mass fluctuations (variance) is
\begin{equation}\label{eq17}
\left\langle {(\delta M/M)^2} \right\rangle \equiv \sigma ^2(M)=\int 
{d^3kW_T^2 (kR)P(k,z),} 
\end{equation}
where W is the Fourier transform of the top-hat window function
\begin{equation}\label{eq18}
\begin{array}{l}
 \mbox{W(}{\rm {\bf x}}\mbox{)}=\left\{ {{\begin{array}{*{20}c}
 {3/4\pi R^3,\mbox{ }\left| {\rm {\bf x}} \right|<R} \hfill \\
 {0,\mbox{ }\left| {\rm {\bf x}} \right|\ge R} \hfill \\
\end{array} }} \right. \\ 
 \to W_T (kR)=3\left[ {\sin (kR)/kR-\cos (kR)} \right]/(kR)^2. \\ 
 \end{array}
\end{equation}
If we approximate P(k) locally with a power-law $P(k,z)=D^2(z)k^m$, where D 
is the linear growth factor, then $\sigma ^2(M)\propto D^2R^{-(3+m)}\propto 
D^2M^{-(3+m)/3}.$ From this we see that the RMS fluctuations are a 
decreasing function of M. At very small mass scales, m$\rightarrow -3$, and 
the fluctuations asymptote to a constant value. We now define the nonlinear 
mass scale by setting $\sigma $(M$_{nl})$=1. We get that (\cite{white94})
\begin{equation}\label{eq19}
M_{nl} (z)\propto D(z)^{6/(3+m)}\mbox{ (}\propto 
\mbox{(1}+\mbox{z)}^{\mbox{-6/(3}+\mbox{m)}}\mbox{ for EdS).}
\end{equation}
For $m > -3$, the smallest mass scales become nonlinear first. This is the 
origin of hierarchical (``bottom-up'') structure formation (Fig. 6b). 

\begin{figure}
\begin{center}
\includegraphics[width=2.5in]{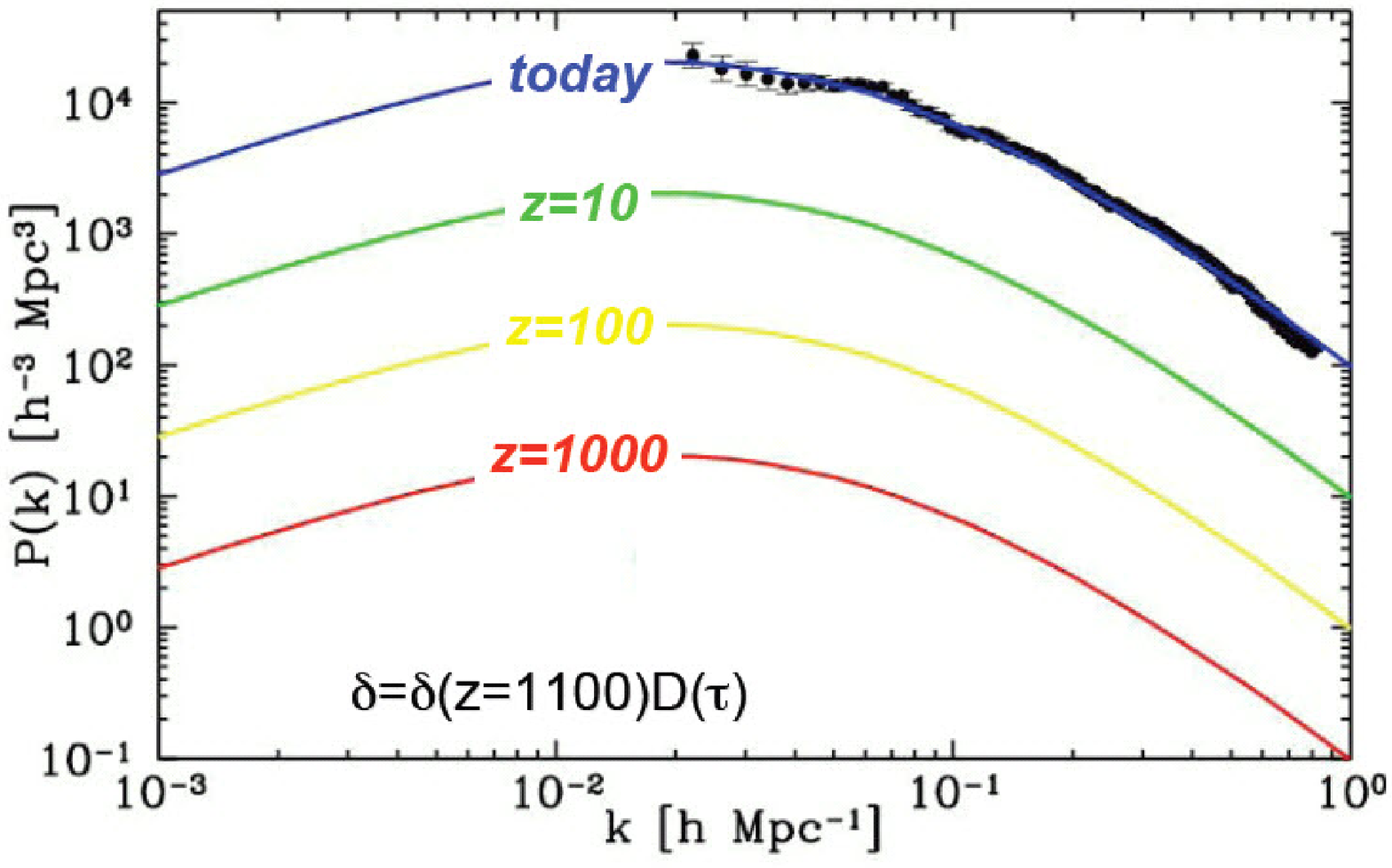}
\includegraphics[width=2.5in]{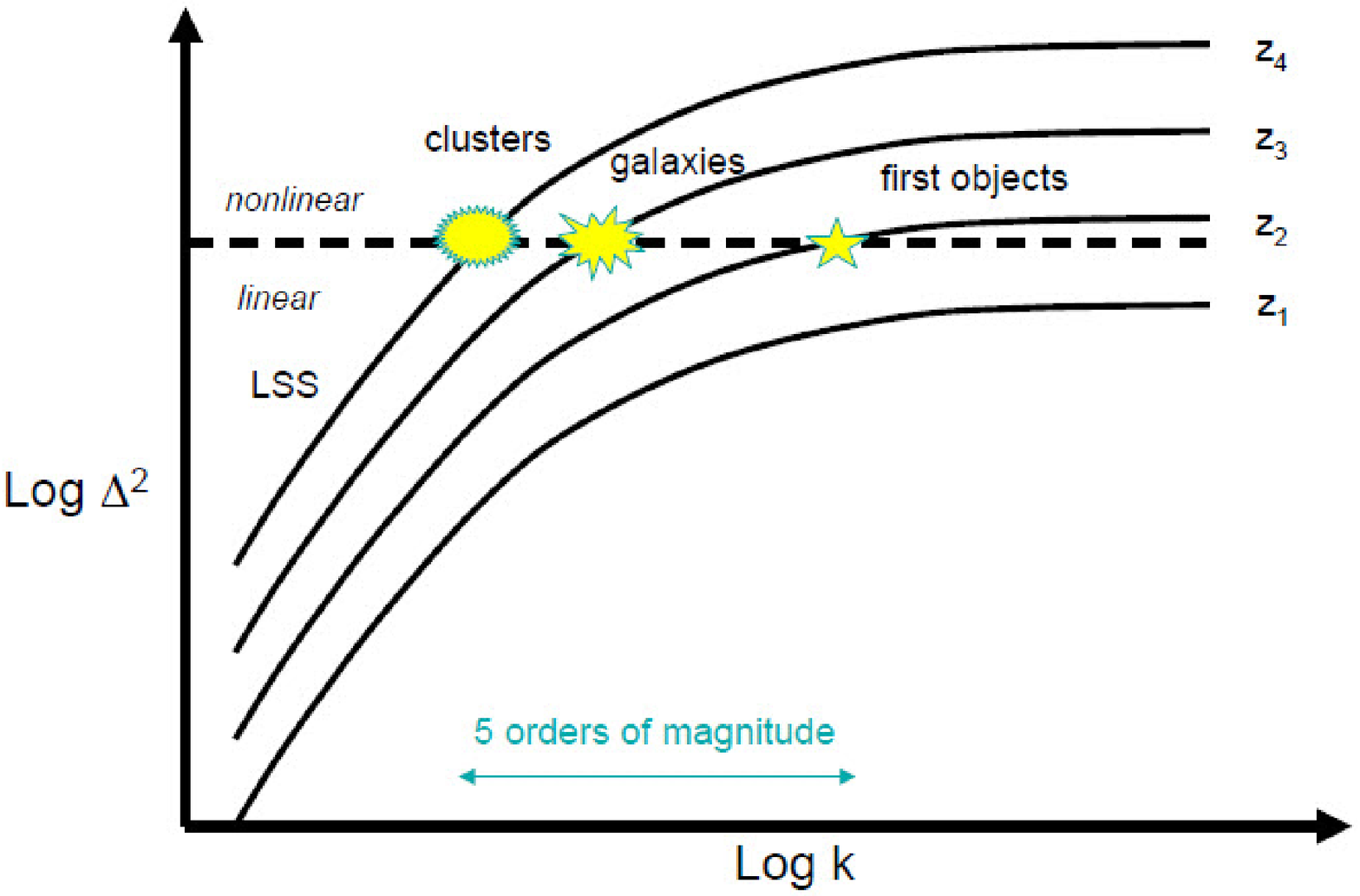}
\end{center}
\caption{Two ways of looking at the growth of post-recombination 
matter fluctuations in the linear regime. Left: 3D matter power spectrum
increases uniformly proportional to the linear growth factor $D(z)$. Measurements
are from SDSS galaxy large scale structure data. Right:
the evolution of the dimensionless power spectrum $\Delta^2 (k)$. Nonlinearity is defined
where $\Delta^2 (k_{nl})=1.$ All scales with $k \geq k_{nl}$ have collapsed into bound
objects, and do so in a ``bottom-up" fashion.}
\label{fig.growth}
\end{figure}

\subsection{Spherical Top-Hat Model}

\begin{figure}[htbp]
\centerline{\includegraphics[width=3in,height=2in]{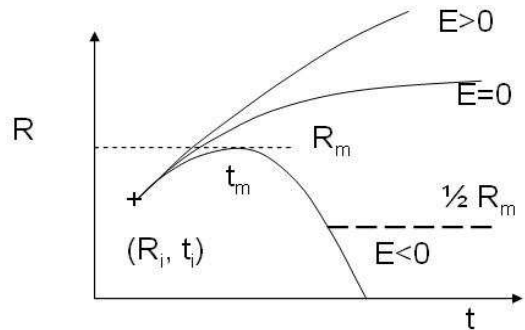}}
\caption{Evolution of a top-hat perturbation in an EdS universe. Depending 
on the E, the first integral of motion, the fluctuation collapses (E$<$0), 
continues to expand (E$>$0), or asymptotically reaches it maximum radius 
(E=0). Virialization occurs when the fluctuation has collapsed to half its 
turnaround radius.}
\label{fig5}
\end{figure}

We now ask what happens when a spherical volume of mass M and radius R 
exceeds the nonlinear mass scale. The simplest analytic model of the 
nonlinear evolution of a discrete perturbation is called the spherical 
top-hat model. In it, one imagines as spherical perturbation of radius $R$
and some constant overdensity $\bar {\delta }=3M/4\pi \bar{\rho} R^3$ in an Einstein-de Sitter 
(EdS) universe. By Birkhoff's theorem the equation of motion for R is 
\begin{equation}\label{eq20}
\frac{d^2R}{dt^2}=-\frac{GM}{R^2}=-\frac{4\pi G}{3}\bar {\rho }(1+\bar 
{\delta })R
\end{equation}
whereas the background universe expands according to Eq. \ref{eq6}
\begin{equation}\label{eq21}
\frac{d^2a}{dt^2}=-\frac{4\pi G}{3}\bar {\rho }a.
\end{equation}

Comparing these two equations, we see that the perturbation evolves like a 
universe of a different mean density, but with the same initial expansion 
rate. Integrating Eq. \ref{eq20} once with respect to time gives us the first 
integral of motion:
\begin{equation}\label{eq22}
\frac{1}{2}\left( {\frac{dR}{dt}} \right)^2-\frac{GM}{R}=E,
\end{equation}
where E is the total energy of the perturbation. If E$<$0, the perturbation 
is bound, and obeys
\begin{equation}\label{eq23}
\frac{R}{R_m}=\frac{(1-cos \theta)}{2}, ~~~\frac{t}{t_m}=\frac{(\theta-sin\theta)}{\pi}
\end{equation}
where $R_m$ and $t_m$ are the radius and time of ``turnaround''. At turnaround 
(as $\theta \rightarrow \pi$), the fluctuation reaches its maximum proper 
radius (see Fig. \ref{fig5}). As 
$t\rightarrow 2t_m, R\rightarrow 0$, and we say the fluctuation has collapsed.

A detailed analysis of the evolution of the top-hat perturbation is given in 
Padmanabhan (1993, Ch. 8) for general $\Omega_m$.
Here we merely quote results for an EdS universe.
The mean \textit{linear} overdensity at turnaround; i.e., the value one would predict from the linear growth formula $\delta \sim t^{2/3}$, is 1.063. The actual overdensity at 
turnaround using the nonlinear model is 4.6. This illustrates that nonlinear 
effects set in well before the amplitude of a linear fluctuation reaches 
unity. As R$\rightarrow $0, the nonlinear overdensity becomes infinite. 
However, the linear overdensity at $t=2t_m$ is only 1.686. As the fluctuation 
collapses, other physical processes (pressure, shocks, violent relation) 
become important which 
establish a gravitationally bound object in virial equilibrium before 
infinite density is reached. Within the framework of the spherical top-hat 
model, we say virialization has occurred when the kinetic and gravitational 
energies satisfy virial equilibrium: $\left| U \right|=2K.$ It is easy to 
show from conservation of energy that this occurs when $R=R_m/2$; in other 
words, when the fluctuation has collapsed to half its turnaround radius. The 
nonlinear overdensity at virialization $\Delta _c$
is not infinite since the radius is finite. 
For an EdS universe, $\Delta _c =18\pi ^2\approx 180$. Fitting
formulae for non-EdS models are provided in the next section.

\subsection{Virial Scaling Relations}
The spherical top-hat model can be scaled to perturbations of arbitrary 
mass. Using virial equilibrium arguments, we can predict various physical 
properties of the virialized object. The ones that interest us most are 
those that relate to the observable properties of gas in galaxy clusters, 
such as temperature, X-ray luminosity, and SZ intensity change. Kaiser \cite{kaiser86}
first derived virial scaling relations for clusters in an EdS universe. Here we 
generalize the derivation to non-EdS models of interest. In order to compute 
these scaling laws, we must assume some model for the distribution of matter 
as a function of radius within the virialized object. A top-hat distribution 
with a density $\rho =\Delta _c \bar {\rho }(z)$ is not useful because it is 
not in mechanical equilibrium. More appropriate is the isothermal, 
self-gravitating, equilibrium sphere for the collisionless matter, whose 
density profile is related to the one-dimensional velocity dispersion 
\cite{bt87}
\begin{equation}\label{eq24}
\rho (r)=\frac{\sigma ^2}{2\pi Gr^2}.
\end{equation}
If we define the virial radius r$_{vir}$ to be the radius of a spherical volume 
within which the mean density is $\Delta _{c}$ times the critical density 
at that redshift ($M=4\pi r_{vir}^3 \rho _{crit} \Delta _c /3)$, then there 
is a relation between the virial mass M and $\sigma $:
\begin{equation}
\label{eq25}
\sigma =M^{1/3}[H^2(z)\Delta _c G^2/16]^{1/6}\approx 476f_\sigma \left( 
{\frac{M}{10^{15}M_\odot }} \right)^{1/3}(h^2\Delta _c E^2)^{1/6}\mbox{ km 
s}^{\mbox{-1}}.
\end{equation}
Here we have introduced a normalization factor $f_{\sigma}$ which will be used to 
match the normailization from simulations. The redshift dependent Hubble 
parameter can be written as $H(z)=100hE(z)\mbox{ km s}^{-1}$ with the 
function $E^2(z)=\Omega _m (1+z)^3+\Omega _k (1+z)^2+\Omega _\Lambda $, 
where the $\Omega$'s have been previously defined. 

The value of $\Delta_c$ is taken from the spherical top-hat model, and is 18$\pi 
^{2}$ for the critical EdS model, but has a dependence on cosmology 
through the parameter $\Omega (z)=\Omega _m (1+z)^3/E^2(z).$ Bryan and Norman 
(1998) provided fitting formulae for $\Delta_c$ for the critical  for both open universe models and flat, lambda-dominated models
\begin{equation}\label{eq26}
\Delta _c =18\pi ^2+82x-39x^2\mbox{ for }\Omega _k =0,\mbox{ }\Delta _c 
=18\pi ^2+60x-32x^2\mbox{ for }\Omega _\Lambda =0
\end{equation}
where x=$\Omega $(z)-1. 

If the distribution of the baryonic gas is also isothermal, we can define a 
ratio of the ``temperature'' of the collisionless material ($T_\sigma =\mu 
m_p \sigma ^2/k)$ to the gas temperature:
\begin{equation}
\label{eq27}
\beta =\frac{\mu m_p \sigma ^2}{kT}
\end{equation}
Given equations (\ref{eq26}) and (\ref{eq27}), the relation between temperature and mass is then
\begin{equation}
\label{eq28}
kT=\frac{GM^{2/3}\mu m_p }{2\beta }\left[ {\frac{H^2(z)\Delta _c }{2G}} 
\right]^{1/3}\approx 1.39f_T \left( {\frac{M}{10^{15}M_\odot }} 
\right)^{2/3}(h^2\Delta _c E^2)^{1/3}\mbox{ keV,}
\end{equation}
where in the last expression we have added the normalization factor f$_{T}$ 
and set $\beta $=1.

The scaling behavior for the object's X-ray luminosity is easily computed by 
assuming bolometric bremsstrahlung emission and ignoring the temperature 
dependence of the Gaunt factor: $L_{bol} \propto 
\int {\rho ^2} T^{1/2}dV\propto M_b \rho T^{1/2}.$ where M$_{b}$ is the 
baryonic mass of the cluster. This is infinite for an isothermal density 
distribution, since $\rho $ is singular. Observationally and 
computationally, it is found that the baryon distribution rolls over to a 
constant density core at small radius. A procedure is described in Bryan and 
Norman (1998) which yields a finite luminosity:
\begin{equation}
\label{eq29}
L_{bol} =1.3\times 10^{45}\left( {\frac{M}{10^{15}M_\odot }} 
\right)^{4/3}(h^2\Delta _c E^2)^{7/6}\mbox{ }\left( {\frac{\Omega _b 
}{\Omega _m }} \right)^2\mbox{ erg s}^{-1}.
\end{equation}
Eliminating M in favor of T in Eq. \ref{eq29} we get
\begin{equation}
\label{eq30}
L_{bol} =6.8\times 10^{44}\left( {\frac{kT/f_T }{1.0\mbox{ keV}}} 
\right)^2(h^2\Delta _c E^2)^{1/2}\mbox{ }\left( {\frac{\Omega _b }{\Omega _m 
}} \right)^2\mbox{ erg s}^{-1}.
\end{equation}
The scaling of the SZ ``luminosity'' is likewise easily computed. If we 
define L$_{SZ}$ as the integrated SZ intensity change: $L_{SZ} =\int {dA\int {n_e 
\sigma _T } } \left( {\frac{kT}{m_e c^2}} \right)dl\propto M_b T$, then
\begin{equation}\label{eq30a}
L_{SZ} =\frac{GM^{5/3}\sigma _T }{2\beta m_e c^2}\left[ {\frac{H^2(z)\Delta 
_c }{2G}} \right]^{1/3}\left( {\frac{\Omega _b }{\Omega _m }} \right).
\end{equation}
We note that cosmology enters these relations only with the combination of 
parameters $h^2\Delta _c E^2$, which comes from the relation between the 
cluster's mass and the mean density of the universe at redshift z. The 
redshift variation comes mostly from E(z), which is equal to (1+z)$^{3/2}$ 
for an EdS universe. 

\subsection{Statistics of hierarchical clustering: Press-Schechter theory}
Now that we have a simple model for the nonlinear evolution of a spherical 
density fluctuation and its observable properties as a function of its 
virial mass, we would like to estimate the number of virialized objects of 
mass M as a function of redshift given the matter power spectrum. This is 
the key to using surveys of galaxy clusters as cosmological probes. While 
large scale numerical simulations can and have been used for this purpose 
(see below), we review a powerful analytic approach by Press and Schechter 
\cite{ps74} which turns out to be remarkably close to the numerical results. The basic 
idea is to imagine smoothing the cosmological density field at any epoch z 
on a scale R such that the mass scale of virialized objects of interest 
satisfies $M=\frac{4\pi }{3}\bar {\rho }(z)R^3.$ Because the density field 
(both smoothed and unsmoothed) is a Gaussian random field, the probability 
that the mean overdensity in spheres of radius R exceeds a critical 
overdensity $\delta _{c}$ is
\begin{equation}\label{eq31}
p(R,z)=\frac{2}{\sqrt {2\pi } \sigma (R,z)}\int\limits_{\delta _c }^\infty 
{d\delta } \exp \left( {-\frac{\delta ^2}{2\sigma ^2(R,z)}} \right)
\end{equation}
where $\sigma(R,z)$ is the RMS density 
variation in spheres of radius R as discussed above. 
Press and Schechter suggested that this probability be identified with the 
fraction of particles which are part of a nonlinear lump with mass exceeding 
M if we take $\delta _c =1.686,$ the linear overdensity at virialization. 
This assumption has been tested against numerical simulations and found to 
be quite good \cite{wef93} (however, see below). The fraction of the 
volume collapsed into objects with mass between $M$ and $M+dM$ is given by 
$(dp/dM)dM$. Multiply this by the average number density of such objects 
$\rho _m /M$ to get the number density of collapsed objects between 
$M$ and $M+dM$:
\begin{equation}\label{eq32}
dn(M,z)=-\frac{\bar {\rho }}{M}\frac{dp(M(R),z)}{dM}dM.
\end{equation}
The minus sign appears here because p is a decreasing function of M. 
Carrying out the derivative using the fact that $dM/dR=3M/R,$ we get
\begin{equation}\label{eq33}
\frac{dn(M,z)}{dM}=\sqrt {\frac{2}{\pi }} \frac{\bar {\rho }\delta _c 
}{3M^2\sigma }e^{-\delta _c^2 /2\sigma ^2}\left[ {-\frac{d\ln \sigma }{d\ln 
R}} \right].
\end{equation}
The term is square brackets is related to the logarithmic slope of the power 
spectrum, which on the mass scale of galaxy clusters is close to unity. Eq. 
\ref{eq33} is called the \textit{halo mass function}, and it has the form of a 
power law multiplied by an exponential. To make this more explicit, approximate the power spectrum on scales of interest as a power law as we have done above. Substituting the 
scaling relations for $\sigma $ in Eq. \ref{eq33} one gets the result \cite{white94}
\begin{equation}\label{eq34}
\frac{dn}{dM}=\left( {\frac{2}{\pi }} \right)^{1/2}\frac{\bar {\rho 
}}{M^2}\left( {1+\frac{m}{3}} \right)\left[ {\frac{M}{M_{nl} (z)}} 
\right]^{\frac{m-3}{6}}\exp \left[ {-\left( {\frac{M}{M_{nl} (z)}} 
\right)^{\frac{3+m}{3}}/2} \right].
\end{equation}
Here, $M_{nl} (z)$ is the nonlinear mass scale. To be more consistent with 
the spherical top-hat model, it satisfies the relation $\sigma (M_{nl} 
,z)=\delta _c $; i.e., those fluctuations in the smoothed density field that 
have reached the linear overdensity for which the spherical top-hat model 
predicts virialization.

\begin{figure}
\begin{center}
\includegraphics[width=\textwidth]{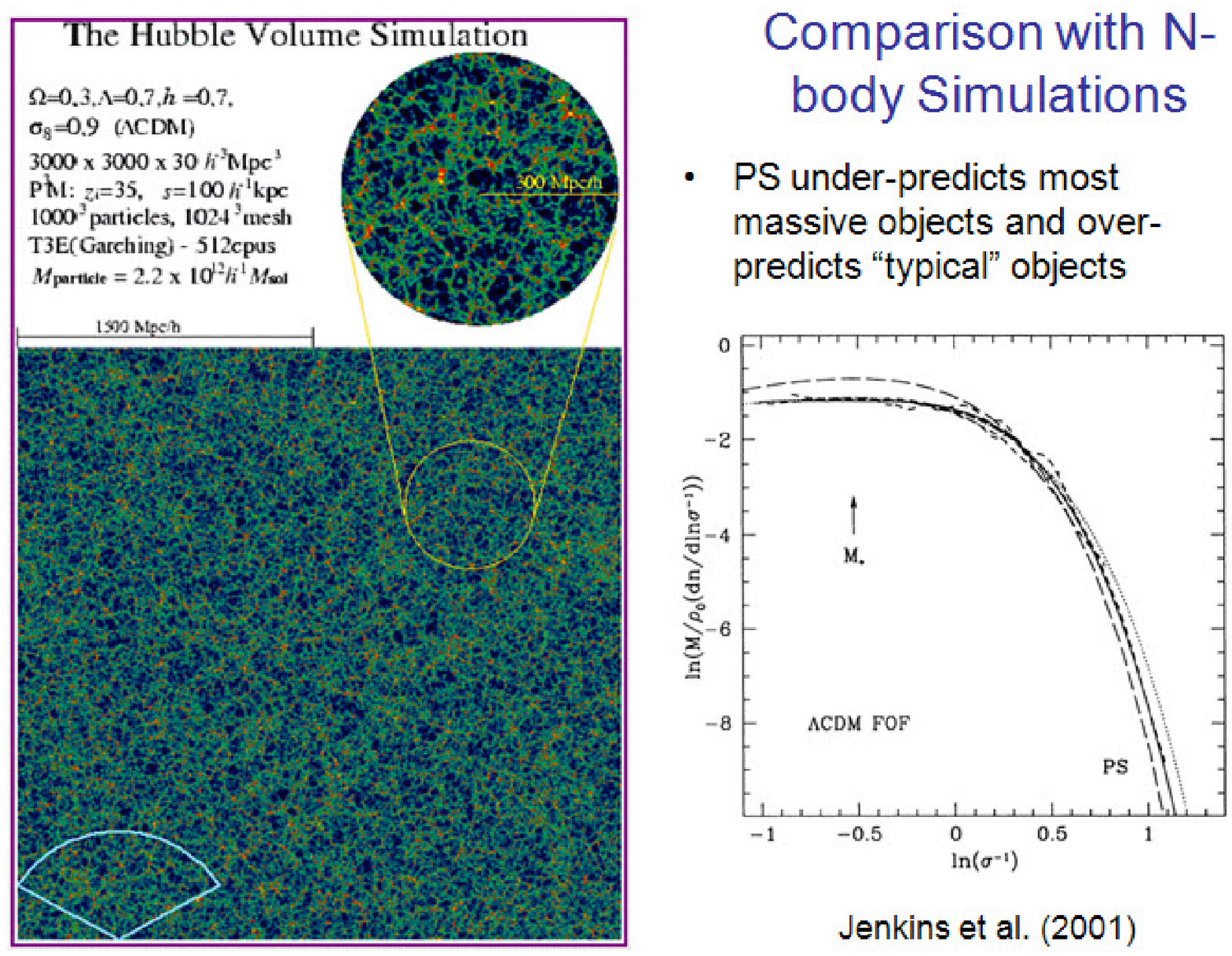}
\end{center}
\caption{The Hubble Volume Simulation \cite{HubbleVolume}. 
Left: slice through the dark matter density
field. Yellow/orange peaks correspond to galaxy clusters. Right: Halo multiplicity function, as measured from the simulation (solid lines), with Press-Schechter 
prediction superposed (dashed line). From \cite{Jenkins01}.}
\label{fig.hubblevolume}
\end{figure}

\subsection{Validating the Halo Mass Function using N-body Simulations}
Below we discuss how one goes about numerically simulating the nonlinear evolution of
the density fluctuations described by the $\Lambda CDM$ power spectrum. Here
we simply mention two works which made detailed comparisons of the PS formula
with halo populations found in dark matter N-body simulations.
The first is by Jenkins et al. (2001)\cite{Jenkins01} who analyzed the results
of the ``Hubble Volume" simulation--a simulation of dark matter clustering carried
out in a cubic volume 3 Gpc/h on a side with $1024^3$ dark matter particles (Fig. \ref{fig.hubblevolume}). 
This yields a dark matter particle mass of $2.2 \times 10^{12}
M_{\odot}$, implying that a galaxy cluster halo would typically contain $100-1000$ particles. The relatively poor mass resolution is offset by the very large volume, 
which permits exploring the cluster mass function across a broad range of masses
including the very high mass end. Fig. \ref{fig.hubblevolume} shows a slice through
the simulation volume on which the dark matter density field is plotted. 
Jenkins et al. (2001) identified dark matter halos 
using the friends-of-friends algorithm \cite{Davis85} 
and found that while the PS formula gives a good approximation
to the numerical data, it underpredicts the number of rare, massive
objects, and overpredicts the number of ``typical" objects as shown in the right
panel of Fig. \ref{fig.hubblevolume}. 

Warren et al. (2006)\cite{Warren06} were interested in testing the validity of the PS formula
over a wider range of mass scales than can be obtained from a single simulation.
They simulated 16 boxes of different physical size but the same number 
of DM particles ($1024^3$) nested in such a way that together they define a
composite halo mass function covering 5 orders of magnitude in mass scale. They derived a fitting formula for the composite halo mass function for the WMAP3 concordance 
cosmological parameters by assuming a parameterized form for the {\em halo multiplicity
function} of:
\begin{equation}
f(\sigma) = A(\sigma^{-a}+b)\exp^{-c/\sigma^2}
\end{equation}
where the multiplicity function $f(\sigma)$ is related to the mass function $n(M)$ via
\begin{equation}
f(\sigma) = \frac{M}{\bar{\rho}} \frac{dn}{d ln \sigma^{-1}}
\end{equation}
where $A, a, b$ and $c$ come from the fitting procedure, and are documented in \cite{Warren06}. 

\begin{figure}[htbp]
\begin{center}
\includegraphics[width=4.0in]{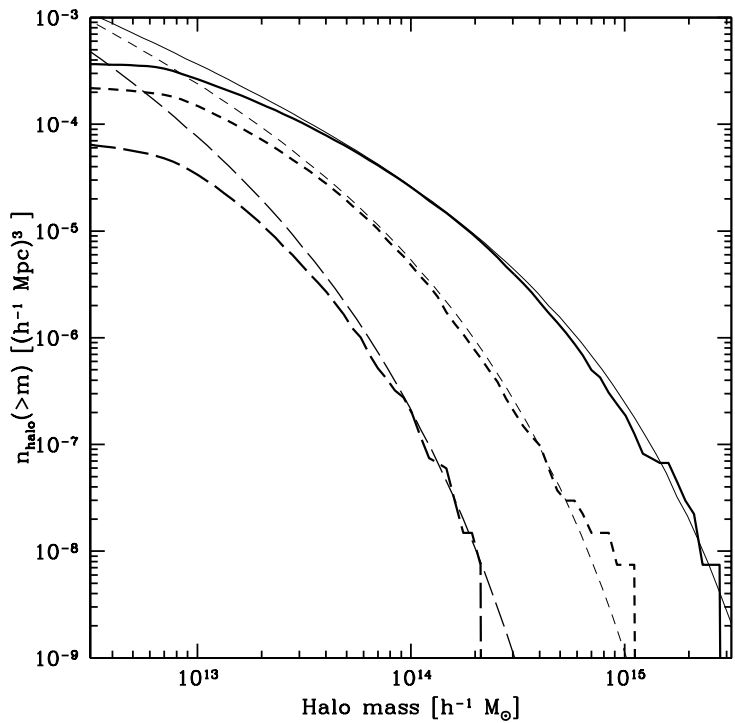}
\end{center}
\caption{Cumulative number of galaxy clusters exceeding mass $m$ for three
redshifts. Solid lines: $z=0.1$, short-dashed lines $z=1$, long-dashed lines $z=2$.
The thick lines are from the simulation shown in Fig. 1, while the thin lines
are predictions using the Warren \cite{Warren06} fitting function. Note the rapid
redshift evolution of the number of massive clusters. 
The departure of the simulation from the predictions at the low mass end are due to finite resolution effects. From \cite{Hallman07}.}
\label{fig.massfn}
\end{figure}

To illustrate one application of this formula, 
Fig. \ref{fig.massfn} shows the cluster halo mass function computed using the
Warren et al. (2006) fit for three different redshifts for the cosmological parameters
adopted in the simulation shown in Fig. 1. Overplotted on the semi-analytic predictions
are the halo mass functions obtained from the simulation itself. The departure of 
the simulation from
the predictions at the low mass end are due to finite resolution effects discussed 
in Sec. 4 below.  

\subsection{Application to galaxy clusters}

\begin{figure}[htbp]
\includegraphics[width=2.5in]{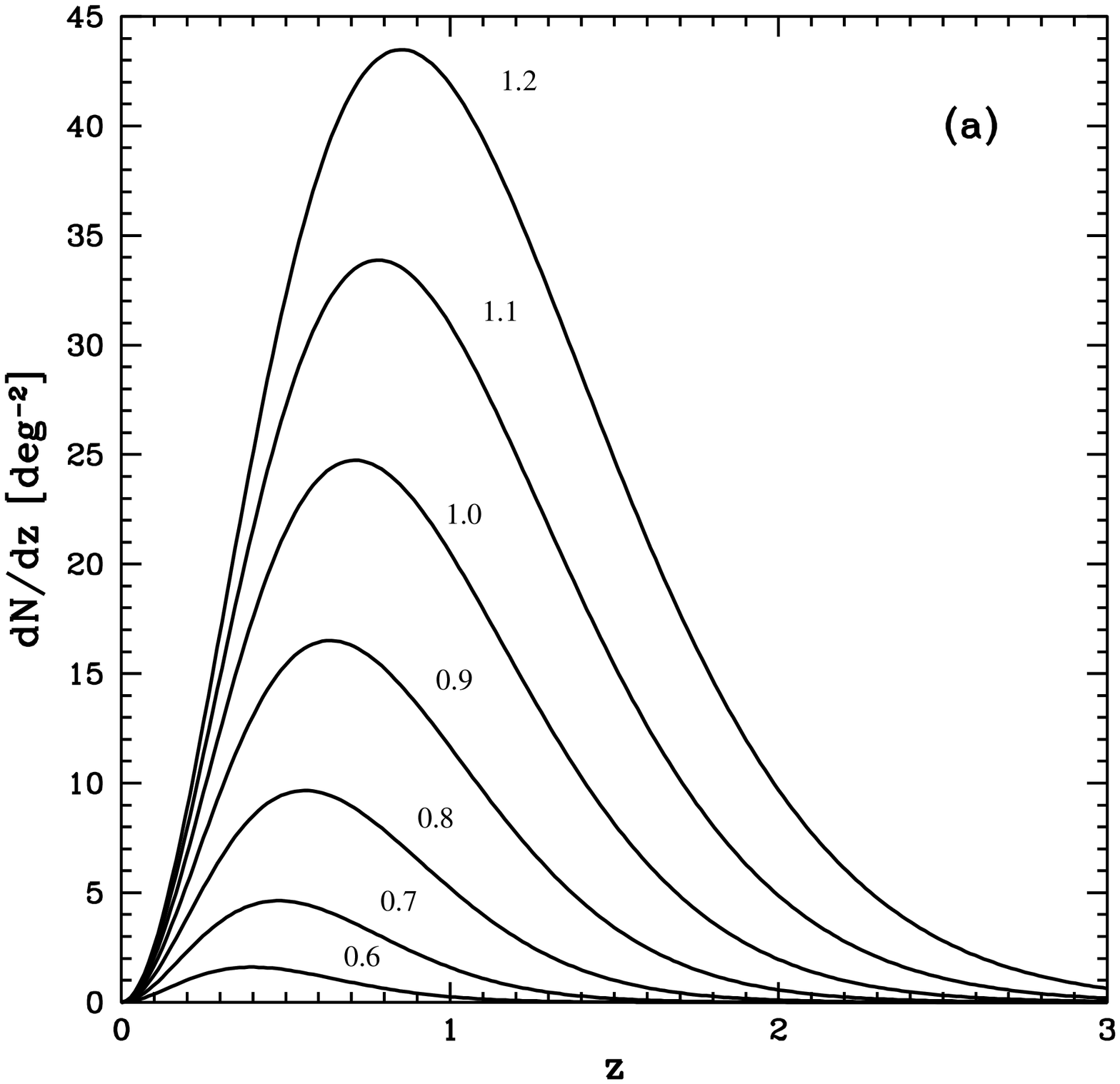}
\includegraphics[width=2.5in]{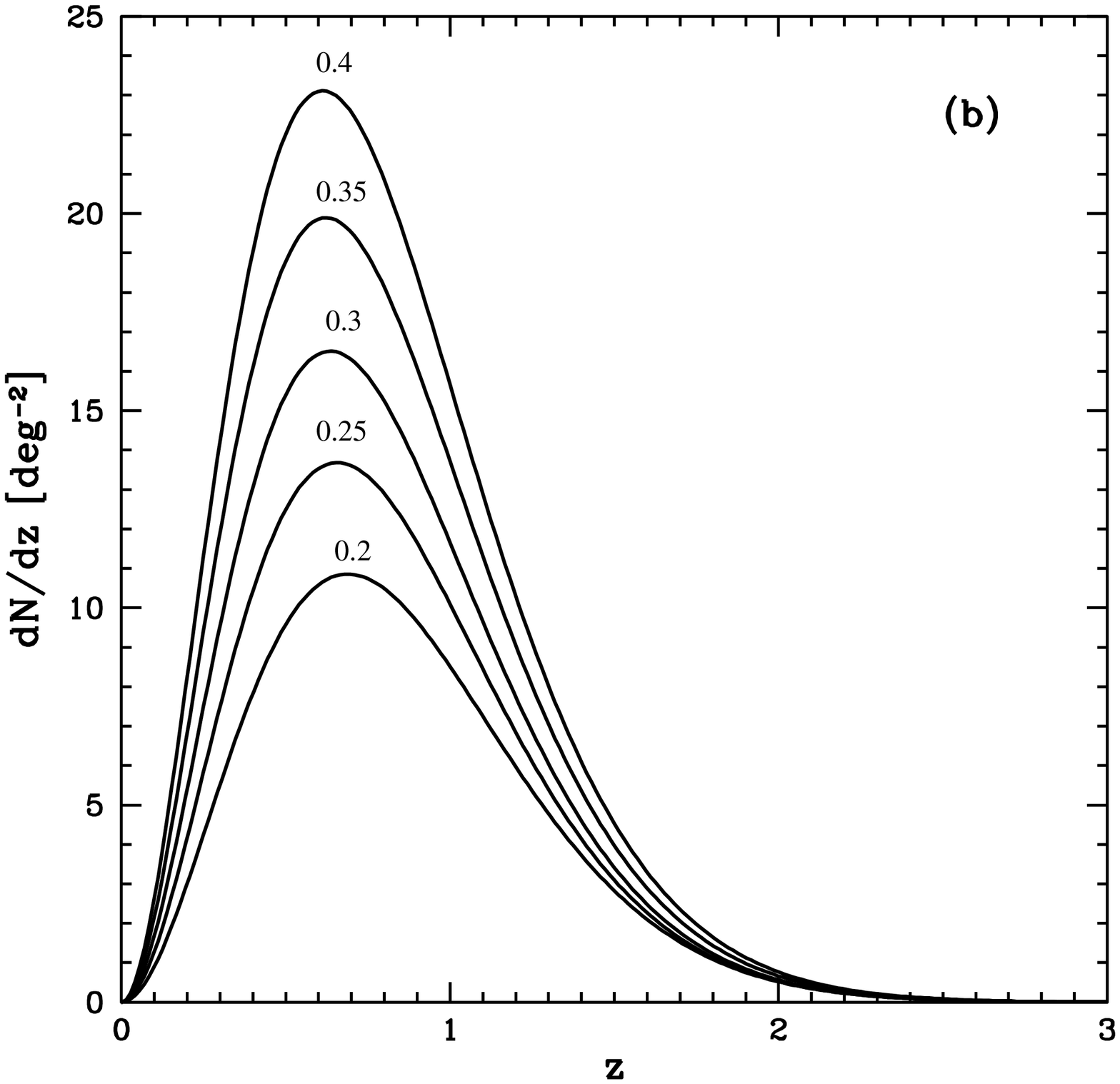}
\caption{Predicted aerial density of galaxy clusters for the WMAP3
concordance cosmological model. Left: varying the amplitude of the 
matter power spectrum $\sigma_8$ holding all other parameters fixed; Right:
varying the matter density $\Omega_m$ holding all other parameters fixed.
From \cite{Hallman07}.}
\label{fig.clusters}
\end{figure}

The aerial density of galaxy clusters can be calculated by multiplying the redshift-dependent halo mass function using the Warren fitting formula with the redshift-dependent differential volume element for one square degree on the sky. Fig. \ref{fig.clusters}a
shows the result varying the amplitude of the 
matter power spectrum $\sigma_8$ holding all other cosmological parameters fixed,
while Fig. \ref{fig.clusters}b shows the result varying the matter density $\Omega_m$ holding all other cosmological parameters fixed. In general we see a rapid rise in the number of clusters with increasing redshift due to the increasing volume element. However, the space density of clusters declines rapidly with redshift (see Fig. \ref{fig.massfn}), and thus the aerial density peaks at $z \approx 1$ and then declines rapidly toward higher redshift. 

From these curves and the virial scaling relations given above
it is easy to predict the expected
number of clusters of a given X-ray temperature, X-ray luminosity, or SZ luminosity
as a function of redshift \cite{ecf96,BN98}, bearing in mind that real clusters
may not perfectly obey the virial scaling relations. In fact they don't as discussed
in my 2004 Varenna lectures \cite{Norman04} and by Borgani in these proceedings.  

\section{Numerical simulations of gas in galaxy clusters}

The central task is for a given cosmological model, calculate the formation 
and evolution of a population of clusters from which synthetic X-ray and SZ 
catalogs can be derived. These can be used to calibrate simpler analytic 
models, as well as to build synthetic surveys (mock catalogs) which can be 
used to assess instrumental effects and survey biases. One would like to 
directly simulate $n(M,z), n(L_x,z), n(T,z), n(Y,z)$ from the governing 
equations for collisionless and collisional matter in an expanding universe. 
Clearly, the quality of these statistical predictions relies on the ability to 
adequately resolve the internal structure and thermodynamical evolution of 
the ICM. 

Since X-ray emission and the SZE are both consequences 
of hot plasma bound in the cluster's gravitational potential well, the 
requirements to faithfully simulate X-ray clusters and SZ clusters are 
essentially the same. Numerical progress can be characterized as a quest for 
higher resolution and essential baryonic physics. In this section I describe 
the technical challenges involved and the numerical methods that have been 
developed to overcome them. I then discuss the effects of assumed baryonic 
physics on ICM structure. Our point of reference is the non-radiative (so-called 
adiabatic) case, which has been the subject of an extensive code comparison 
\cite{Frenk99}. 

In Norman (2003) \cite{Norman03}
I provided a historical review of the progress that has 
been made in simulating the evolution of gas in galaxy clusters motivated by 
X-ray observations. 
In Norman (2004) \cite{Norman04} I discuss the statistical properties
of simulated galaxy cluster samples and how they depend on assumed baryonic 
physics. The key result of this work is that while $L_x$ is highly sensitive
to input physics and numerical resolution, $L_{SZ}$ is not, and therefore 
potentially a useful proxy for the cluster mass and thereby a cosmological probe.
I discuss recent progress on increasing the physics fidelity of individual
cluster simulations in Sec. 5, and the use of cluster SZ surveys as cosmological
probes in Sec. 6.

\subsection{Dynamic range considerations}

\begin{figure}[htbp]
\includegraphics[width=3in,height=1.7in]{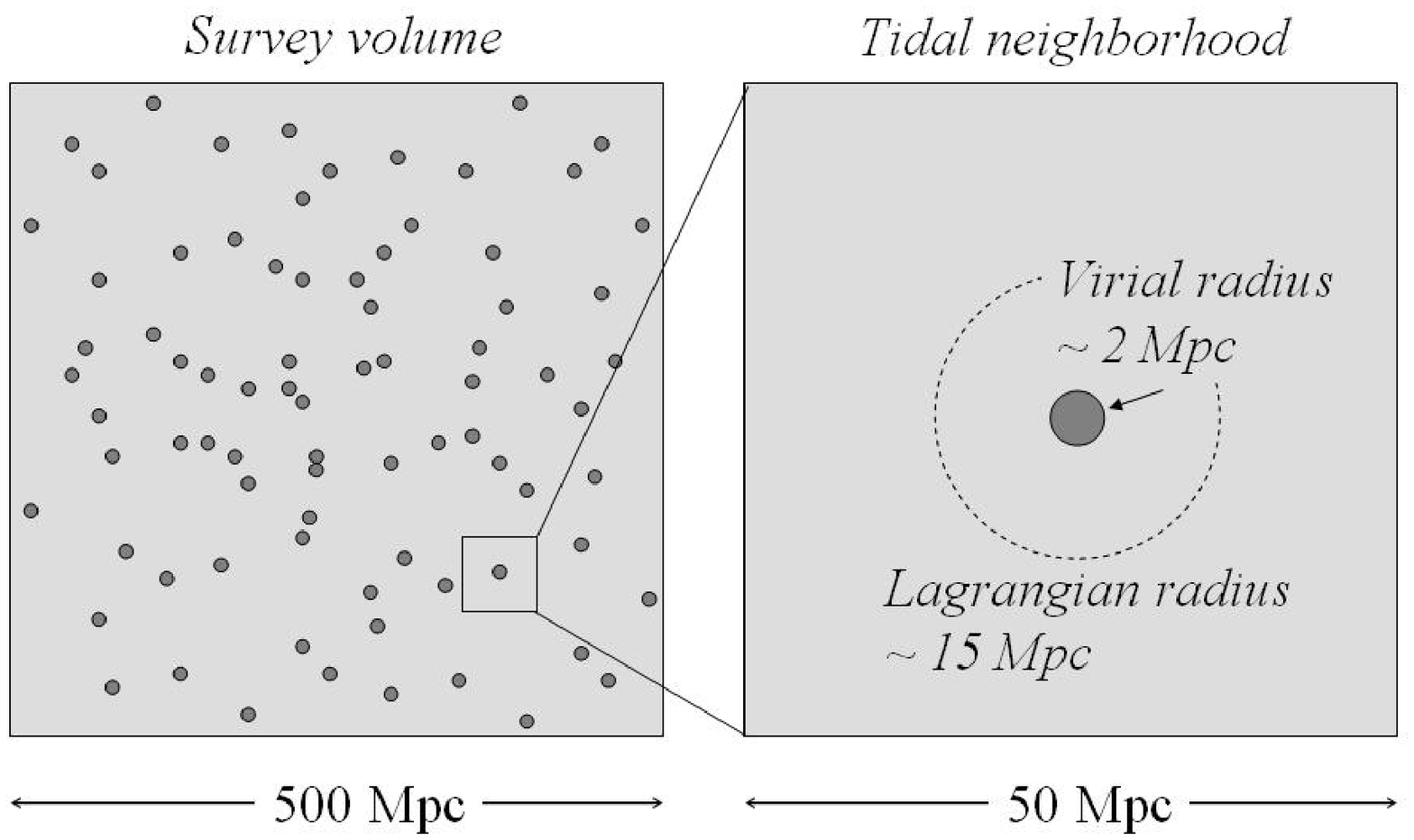}
\includegraphics[width=2.5in,height=1.7in]{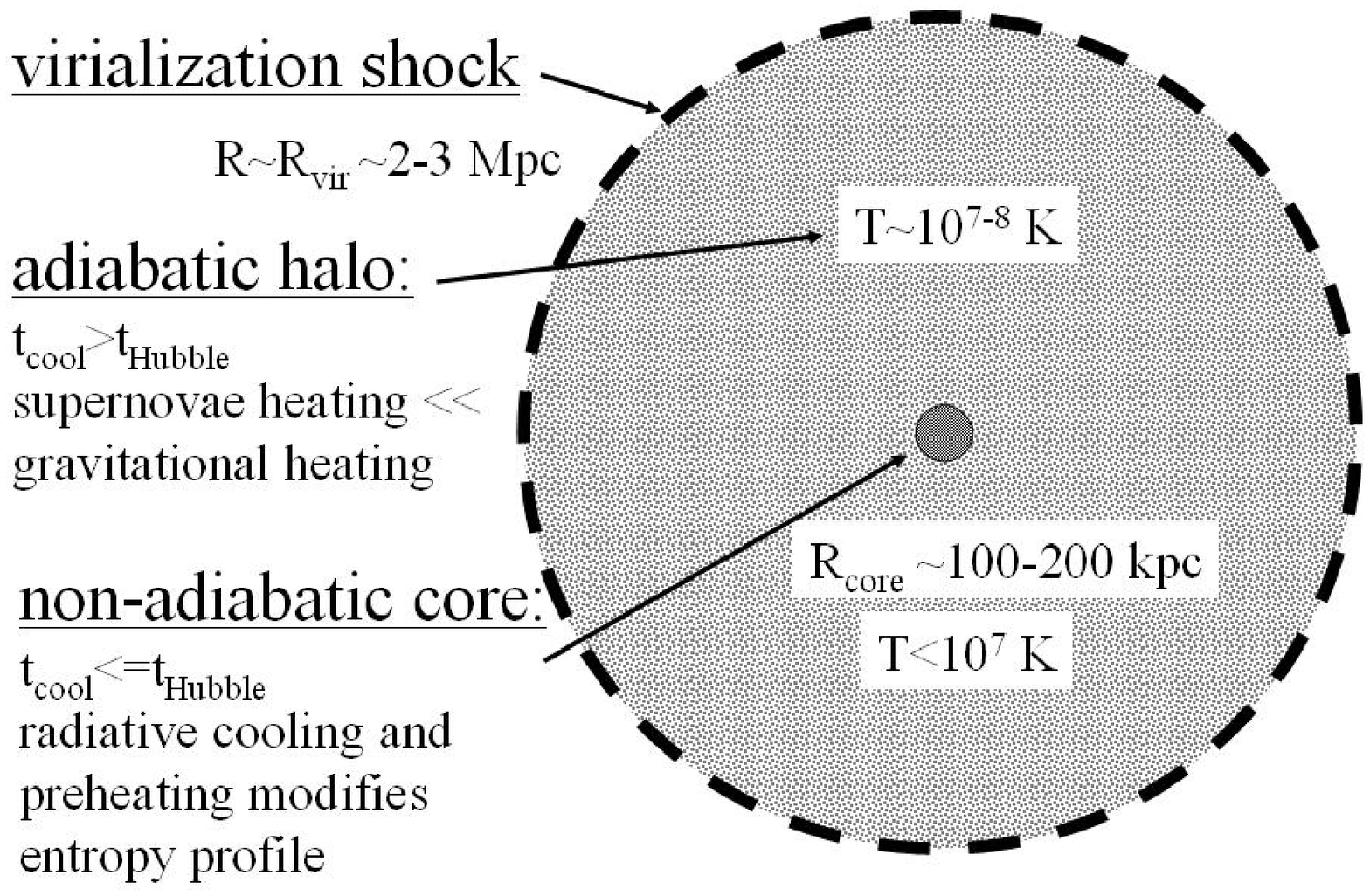}
\caption{Left: A range of length scales of $\sim $250 separates the size of a reasonable survey volume and the virial radius of a rich cluster. 
Right: Simplified structure of the ICM in a massive cluster. A range of length scales of $\sim $20-30 separates the virial radius and the core radius. }
\label{fig7}
\end{figure}

Fig. \ref{fig7} illustrates the dynamic range difficulties encountered with 
simulating a statistical ensemble of galaxy clusters, while at the same time 
resolving their internal structure. Massive clusters are rare at any 
redshift, yet these are the ones most that are most sensitive to cosmology. 
From the cluster mass function (Fig. \ref{fig.massfn}), in order to get adequate 
statistics, one deduces that one must simulate a survey volume many hundreds 
of megaparsecs on a side (Fig. \ref{fig7}, left panel). 
A massive cluster has a virial radius of 
$\sim $2 Mpc. It forms via the collapse of material within a comoving 
Lagrangian volume of $\sim $15 Mpc. However, tidal effects from a larger 
region (50-100 Mpc) are important on the dynamics of cluster formation. The 
internal structure of cluster's ICM is shown in Fig. \ref{fig7}, center panel. 
While clusters are 
not spherical, two important radii are generally used to characterize them: 
the virial radius, which is the approximate location of the virialization 
shock wave that thermalizes infalling gas to 10-100 million K, and the core 
radius, within which the baryon densities plateau and the highest X-ray 
emissions and SZ intensity changes are measured. A typical radius is $\sim $200 
kpc. Within the core, radiative cooling and possibly other physical 
processes are important. Outside the core, cooling times are longer than the 
Hubble time, and the ICM gas is effectively adiabatic. If we wanted to achieve 
a spatial resolution of 1/10 of a core radius everywhere within the survey 
volume, we would need a spatial dynamic range of D=500 Mpc/20 kpc = 25,000. 
The mass dynamic range is more severe. If we want 1 million dark matter 
particles within the virial radius of a $10^{15} M_{\odot}$ cluster, then we would 
need $N_{particle} =M_{box} /M_{particle} =\Omega _m \rho _{crit} 
L^3/10^9\approx 10^{11}$ if they were uniformly distributed in the survey 
volume. 

Two solutions to spatial dynamic range problem have been developed: tree 
codes for gridless N-body methods \cite{KWH96,syw01} 
and adaptive mesh refinement (AMR) for Eulerian particle-mesh/hydrodynamic methods \cite{bn97,Kravtsov97,Teyssier02,OShea04}. 
Both methods increase the spatial 
resolution automatically in collapsing regions as described below. The 
solution to the mass dynamic range problem is the use of multi-mass initial 
conditions in which a hierarchy of particle 
masses is used, with many low mass particles concentrated in the region of 
interest. This approach has most recently used by Springel et al. (2000)
\cite{springel00}, 
who simulated the formation of a galaxy cluster dark matter halo with 
$N=6.9\times 10^6$ dark matter particles, resolving the dark matter halos 
down to the mass scale of the Fornax dwarf spheroidal galaxy. The spatial 
dynamic range achieved in this simulation was $R=2\times 10^5$. Such dynamic 
ranges have not yet been achieved in galaxy cluster simulations with gas.

\begin{figure}
\begin{center}
\includegraphics[width=0.9\textwidth]{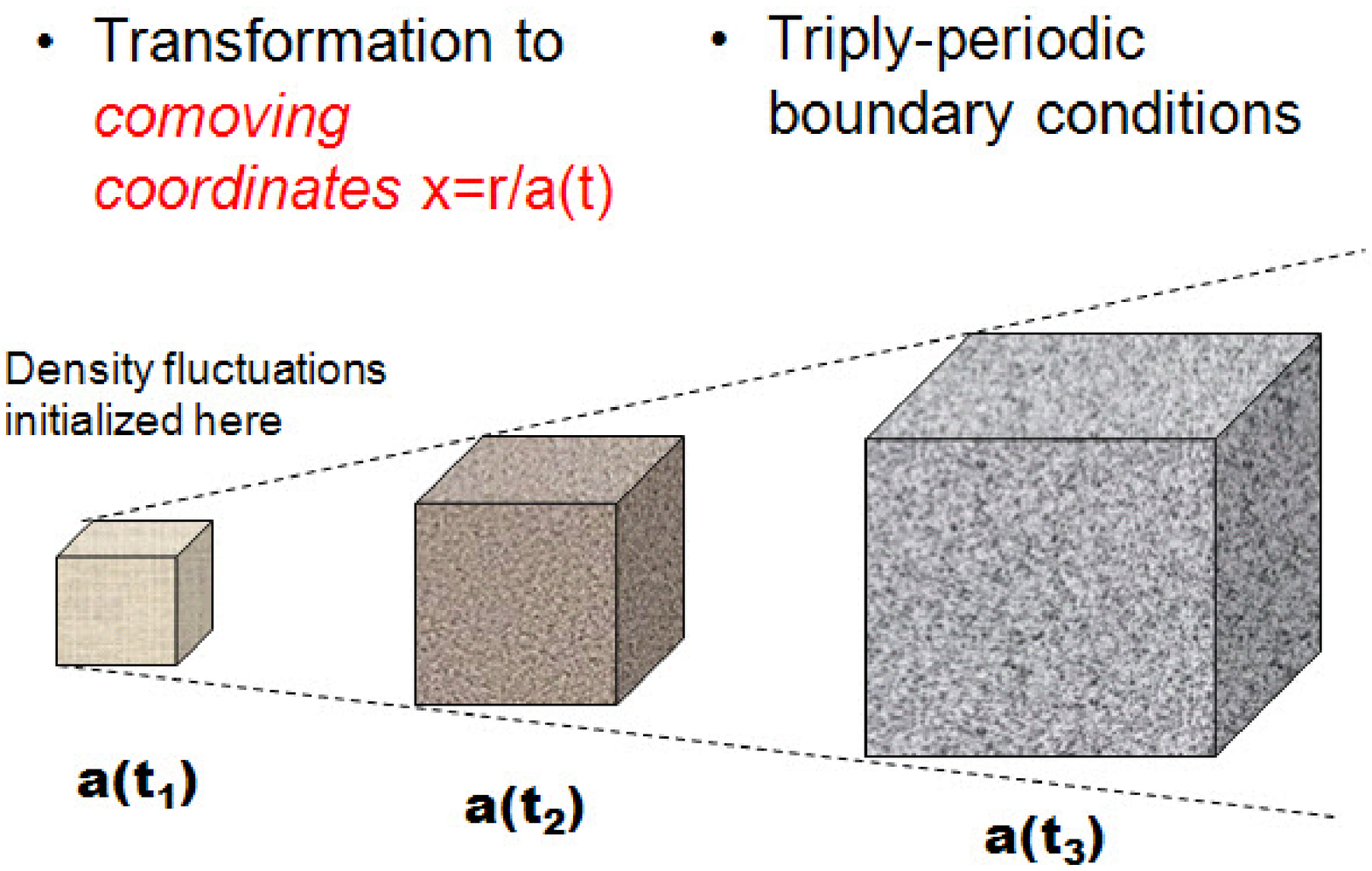}
\end{center}
\caption{Cosmological simulations are generally carried out in a frame of reference
that is comoving with the expanding universe. Initial conditions are generated
from the input power spectrum at a starting redshift and then advanced in time
using equations \ref{eq35} -- \ref{eq37}.}
\label{fig.comoving}
\end{figure}

\subsection{Simulating cluster formation}

Simulations of cosmological structure formation are done in a cubic domain 
which is comoving with the expanding universe (cf. Fig. \ref{fig.comoving}). 
Matter density and velocity 
fluctuations are initialized at the starting redshift chosen such that all 
modes in the volume are still in the linear regime. 
Once initialized, these fluctuations are then evolved to 
z=0 by solving the equations for collisionless N-body dynamics for cold dark 
matter, and the equations of ideal gas dynamics for the baryons in an 
expanding universe. Making the transformation from proper to comoving 
coordinates $\vec {r}=a(t)\vec {x}$, Newton's laws for the collsionless dark 
matter particles become
\begin{equation}
\label{eq35}
\frac{d\vec {x}_{dm} }{dt}=\vec {\upsilon }_{dm} ,\mbox{ }\frac{d\vec 
{\upsilon }_{dm} }{dt}=-2\frac{\dot {a}}{a}\vec {\upsilon }_{dm} 
-\frac{1}{a^2}\nabla _x \phi 
\end{equation}
where $x$ and $v$ are the particle's comoving position and peculiar velocity, 
respectively, and $\phi$ is the comoving gravitational potential that includes 
baryonic and dark matter contributions. The hydrodynamical equations for 
mass, momentum, and energy conservation in an expanding universe in comoving 
coordinates are (\cite{Anninos97})
\begin{equation}
\label{eq36}
\begin{array}{l}
 \frac{\partial \rho _b }{\partial t}+\nabla \cdot (\rho _b \vec {\upsilon 
}_b )+3\frac{\dot {a}}{a}\rho _b =0, \\ 
 \frac{\partial (\rho _b \upsilon _{b,i} )}{\partial t}+\nabla \cdot [(\rho 
_b \upsilon _{b,i} )\vec {\upsilon }_b +5\frac{\dot {a}}{a}\rho _b \upsilon 
_{b,i} =-\frac{1}{a^2}\frac{\partial p}{\partial x_i }-\frac{\rho _b 
}{a^2}\frac{\partial \phi }{\partial x_i }, \\ 
 \frac{\partial e}{\partial t}+\nabla \cdot (e\vec {\upsilon }_b )+p\nabla 
\cdot \vec {\upsilon }_b +3\frac{\dot {a}}{a}e=\Gamma -\Lambda , \\ 
 \end{array}
\end{equation}
where $\rho_b, p$ and $e$, are the baryonic density, pressure and internal energy 
density defined in the proper reference frame, $\vec {\upsilon }_b $ is the 
comoving peculiar baryonic velocity, $a=1/(1+z)$ is the cosmological scale 
factor, and $\Gamma $ and $\Lambda $ are the microphysical heating and 
cooling rates. The baryonic and dark matter components are coupled through 
Poisson's equation for the gravitational potential
\begin{equation}
\label{eq37}
\nabla ^2\phi =4\pi Ga^2(\rho _b +\rho _{dm} -\bar {\rho }(z))
\end{equation}
where $\bar {\rho }(z)=3H_0 \Omega _m (0)/8\pi Ga^3$ is the proper 
background density of the universe.

The cosmological scale factor $a(t)$ is obtained by integrating the 
Friedmann equation (Eq. \ref{eq4}). To complete the specification of the problem we 
need the ideal gas equation of state $p=(\gamma -1)e$, and the gas heating 
and cooling rates. When simulating the ICM, the simplest approximation is to 
assume $\Gamma $ and $\Lambda =0$; i.e., no heating or cooling of 
the gas other than by adiabatic processes and shock heating. 
Such simulations are referred to as 
adiabatic (despite entropy-creating shock waves), and are a reasonable first approximation to real clusters because 
except in the cores of clusters, the radiative cooling time is longer than a 
Hubble time, and gravitational heating is much larger than sources of 
astrophysical heating. However, as discussed in the paper by Cavaliere in 
this volume, there is strong evidence that the gas in cores of clusters has 
evolved non-adiabatically. This is revealed by the entropy profiles observed 
in clusters \cite{Ponman99} which deviate substantially from adiabatic 
predictions. In the simulations presented below, we consider radiative 
cooling due to thermal bremsstrahlung, and mechanical heating due to galaxy 
feedback, details of which are described below.

\begin{figure}
\begin{center}
\includegraphics[width=4.0in]{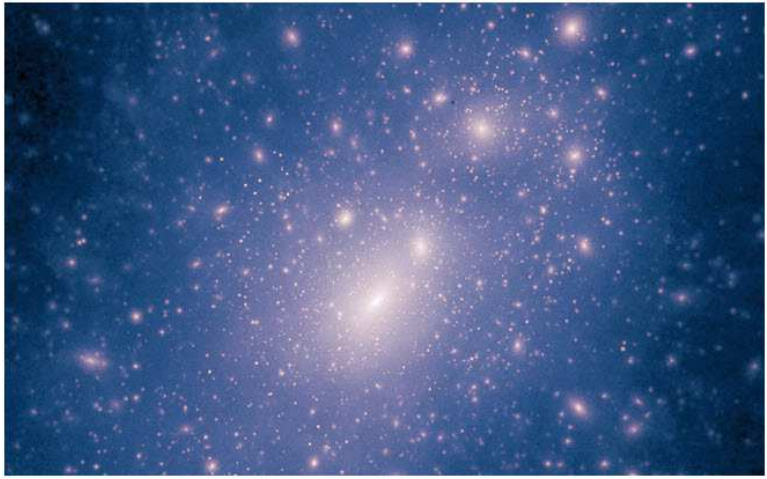}
\end{center}
\caption{Ultra-high resolution N-body simulations of the clustering of dark matter 
reveal a wealth of substructure. From \cite{GADGET2}.}
\label{fig.DM}
\end{figure}

\subsection{Numerical methods overview}

A great deal of literature exists on the gravitational clustering of CDM 
using N-body simulations. A variety of methods have been employed including 
the fast grid-based methods particle-mesh (PM), and 
particle-particle+particle-mesh (P$^{3}$M) \cite{Efstathiou81}, 
spatially adaptive methods such as adaptive P$^{3}$M \cite{Couchman91}, 
adaptive mesh refinement \cite{Kravtsov97}, tree codes 
\cite{BarnesHut86,WarrenSalmon94}, and hybrid methods such as TreePM 
\cite{Xu99}. Because of the large dynamic range required, 
spatially adaptive methods are favored, with Tree and TreePM methods the 
most widely used today. Fig. \ref{fig.DM} shows a high resolution N-body
simulation of the substructure within a dark matter halo performed by
Springel using the GADGET code \cite{GADGET2}.

When gas dynamics is included, only certain 
combinations of hydrodynamics algorithms and collisionless N-body algorithms 
are ``natural''. Dynamic range considerations have led to two principal 
approaches: P$^{3}$MSPH and TreeSPH, which marries a P$^3$M or tree code for 
the dark matter with the Lagrangian smoothed-particle-hydrodynamics (SPH) 
method \cite{Evrard88,KWH96,syw01}, and adaptive mesh refinement (AMR), 
which marries PM with 
Eulerian finite-volume gas dynamics schemes on a spatially adaptive mesh 
\cite{bn97,OShea04,Teyssier02,Kravtsov03}.
Pioneering hydrodynamic simulations 
using non-adaptive Eulerian grids 
\cite{Kang94,Bryan94,BN98}
yielded some important insights about cluster formation and 
statistics, but generally have inadequate resolution to resolve their internal 
structure in large survey volumes. In the following we concentrate on our 
latest results using the AMR code \textit{Enzo} \cite{OShea04}. 
The reader is also 
referred to the paper by Borgani et al. \cite{Borgani04} which presents recent, 
high-resolution results from a large TreeSPH simulation.

\textit{Enzo} is a grid-based hybrid code (hydro + N-body) which uses the 
block-structured AMR algorithm of Berger {\&} Collela \cite{Berger89} to improve 
spatial resolution in regions of large gradients, such as in gravitationally 
collapsing objects. The method is attractive for cosmological applications 
because it: (\ref{eq1}) is spatially- and time-adaptive, (\ref{eq2}) uses accurate and 
well-tested grid-based methods for solving the hydrodynamics equations, and 
(\ref{eq3}) can be well optimized and parallelized. The central idea behind AMR is 
to solve the evolution equations on a grid, adding finer meshes in regions 
that require enhanced resolution. Mesh refinement can be continued to an 
arbitrary level, based on criteria involving any combination of overdensity 
(dark matter and/or baryon), Jeans length, cooling time, etc., enabling us 
to tailor the adaptivity to the problem of interest. The code solves the 
following physics models: collisionless dark matter and star particles, 
using the particle-mesh N-body technique \cite{Hockney88}; gravity, using FFTs on the 
root grid and multigrid relaxation on the subgrids; cosmic expansion; gas 
dynamics, using the piecewise parabolic method (PPM)\cite{Collela84}; 
multispecies 
nonequilibrium ionization and H$_{2}$ chemistry, using backward Euler time 
differencing \cite{Anninos97}; radiative heating and cooling, using subcycled forward 
Euler time differencing 
\cite{Anninos94}; and a parameterized star formation/ feedback 
recipe \cite{Cen92}. At the present time, magnetic fields and radiation transport 
are being installed. \textit{Enzo} is publicly available at
{\textit{http://lca.ucsd.edu/projects/enzo}}.

\subsection{Structure of nonradiative clusters: the Santa Barbara test cluster}

\begin{figure}[htbp]
\includegraphics[width=5in,height=3.33in]{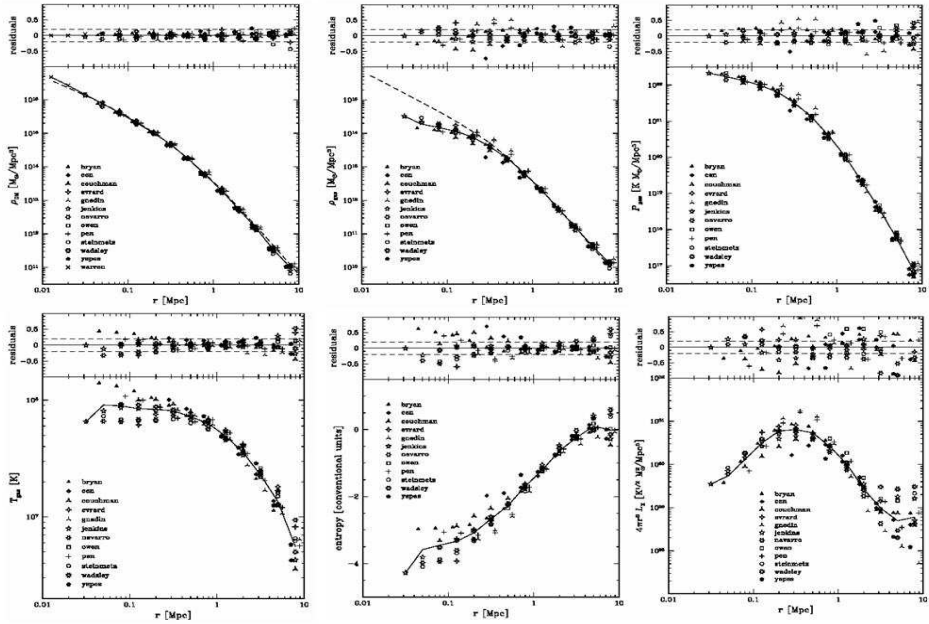}
\caption{The Santa Barbara test cluster. Top row, left to right: profiles
of dark matter density, gas density, and gas pressure. Bottom row, left
to right: profiles of gas temperature, gas entropy, and X-ray emissivity.
Different symbols correspond to different code results. From \cite{Frenk99}.}
\label{fig.SBcluster}
\end{figure}

In Frenk et al. \cite{Frenk99} 12 groups compared the results of a variety of 
hydrodynamic cosmological algorithms on a standard test problem. The test 
problem, called the Santa Barbara cluster, was to simulate the formation 
of a Coma-like cluster in a standard CDM cosmology ($\Omega_m=1$) 
assuming the gas is nonradiative. Groups 
were provided with uniform initial conditions and were asked to carry out a 
``best effort'' computation, and analyze their results at z=0.5 and z=0 for 
a set of specified outputs. These outputs included global integrated quantities, 
radial profiles, and column-integrated images. The simulations varied 
substantially in their spatial and mass resolution owing to algorithmic and 
hardware limitations. Nonetheless, the comparisons brought out which 
predicted quantities were robust, and which were not yet converged. In Fig. 
\ref{fig.SBcluster} we show a few figures from Frenk et al. (1999) which highlight areas of 
agreement (top row) and disagreement (bottom row). 

The top row shows profile of dark matter density, baryon density, and 
pressure for the different codes. All are in quite good agreement for the 
\textit{mechanical structure} of the cluster. The dark matter profile is well described by an NFW profile 
which has a central cusp \cite{NFW96}. The baryon 
density profiles show more dispersion, but all codes agree that the profile 
flattens at small radius, as observed. All codes agree extremely well on the 
gas pressure profile, which is not surprising, since mechanical equilibrium 
is easy to achieve for all methods even with limited resolution. This bodes 
well for the interpretation of SZE observations of clusters, since the 
Compton y parameter is proportional to the projected pressure distribution. 
In section 5 we show results from a statistical ensemble of clusters which 
bear this out. 

The bottom row shows the thermodynamic structure of the cluster, as well as 
the profile of X-ray emissivity. The temperature profiles show a lot of 
scatter within about one-third the virial radius (=2.7 Mpc). 
Systematically, the SPH codes produce nearly isothermal cores, while the 
grid codes produce temperature profiles which continue to rise as 
r$\rightarrow $0. The origin of this discrepancy has not been resolved, but 
improved SPH formulations come closer to reproducing the AMR results
\cite{Ascasibar03}. This discrepancy is reflected in the entropy 
profiles. Again, agreement is good in the outer two-thirds of the cluster, but the 
profiles show a lot of dispersion in the inner one third. Discounting the codes 
with inadequate resolution, one finds the SPH codes produce an entropy 
profile which continues to fall as r$\rightarrow $0, while the grid codes 
show an entropy core, which is more consistent with observations \cite{Ponman99}.
The dispersion in the density and temperature profiles are 
amplified in the X-ray emissivity profile, since $\varepsilon _x \propto 
\rho _b^2 T^{1/2}$. The different codes agree on the integrated X-ray luminosity of 
the cluster only to within a factor of 2. This is primarily because the 
density profile is quite sensitive to resolution in the core; any 
underestimate in the core density due to inadequate resolution is amplified 
by the density squared dependence of the emissivity. This suggests that 
quite high resolution is needed, as well as a good grasp on non-adiabatic 
processes operating in cluster cores, before simulations will be able to 
accurately predict X-ray luminosities. 

\subsection{Effect of additional physics}

\begin{figure}[htbp]
\includegraphics[width=3in,height=3.5in]{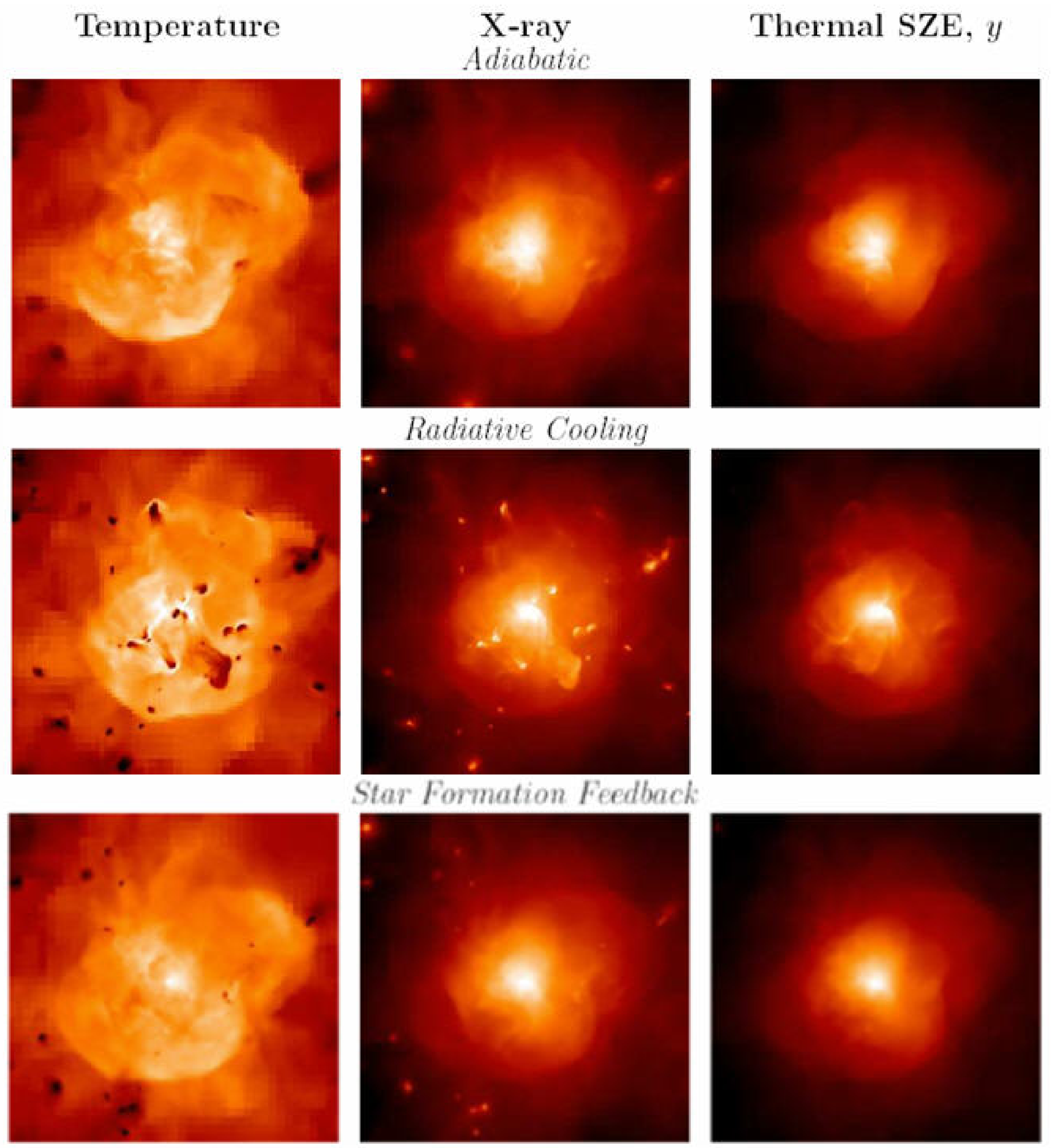}
\includegraphics[width=2.5in,height=2in]{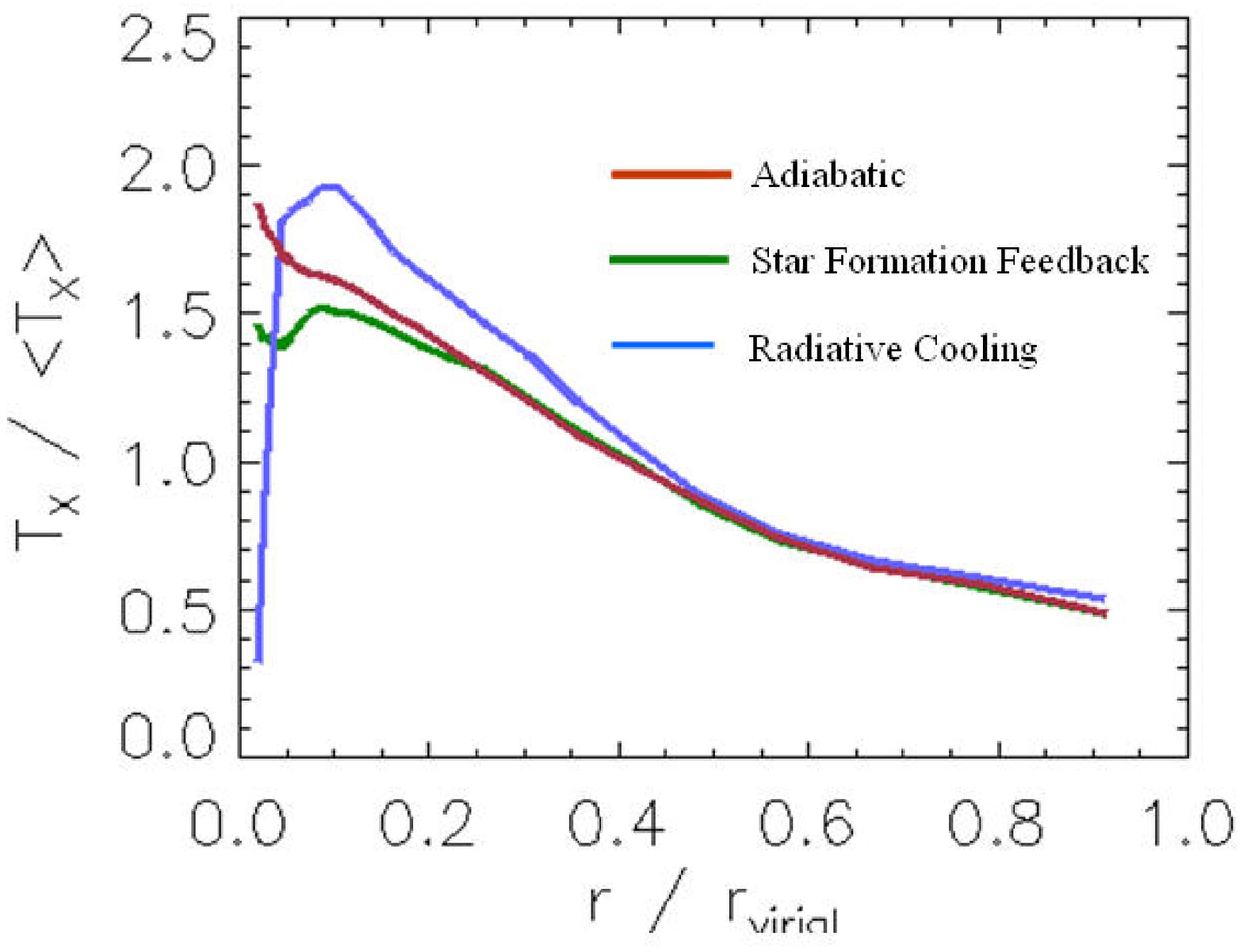}
\caption{Effect of physical processes on simulated galaxy cluster ICM observables.
Left: Columns show X-ray surface brightness, projected temperature, and Compton y-parameter for a $M=2\times 10^{15} M_{\odot}$ cluster assuming different baryonic physics. Field of view is 5 h$^{-1}$ Mpc. Right:
Corresponding spherically averaged radial temperature profiles. From \cite{Motl05}.}
\label{fig10}
\end{figure}

Within r=0.15 r$_{vir}$, Vikhlinin et al. \cite{Vikhlinin04}
found large variation in 
temperature profiles, but in all cases the gas is cooler than the cluster 
mean. This suggests that radiative cooling is important in cluster cores, 
and possibly other effects as well. It has been long known that $\sim 
60$ percent of nearby, luminous X-ray clusters have central X-ray excesses, 
which has been interpreted as evidence for the presence of a cluster-wide 
cooling flows \cite{Fabian94}. More recently, Ponman et al. \cite{Ponman99}
have used X-ray 
observations to deduce the entropy profiles in galaxy groups and clusters. 
They find an entropy floor in the cores of clusters indicative of extra, 
non-gravitational heating, which they suggest is feedback from galaxy 
formation. It is easy to imagine cooling and heating both may be important 
to the thermodynamic evolution of ICM gas. 

To explore the effects of additional physics on the ICM, we recomputed the 
entire sample of clusters changing the assumed baryonic physics, keeping 
initial conditions the same. Three additional samples of about 100 clusters 
each were simulated: The ``radiative cooling'' sample assumes no additional 
heating, but gas is allowed to cool due to X-ray line and bremsstrahlung 
emission in a 0.3 solar metallicity plasma. The ``star formation'' sample 
uses the same cooling, but additionally cold gas is turned into 
collisionless star particles at a rate $\dot {\rho }_{SF} =\varepsilon _{sf} 
\frac{\rho _b }{\max (\tau _{cool} ,\tau _{dyn} )}$ , where $\varepsilon 
_{sf}$ is the star formation efficiency factor $\sim $0.1, and $\tau 
_{cool}$ and $\tau _{dyn}$ are the local cooling time and freefall time, 
respectively. This locks up cold baryons in a non-X-ray emitting component, 
which has been shown to have an important effect of the entropy profile of 
the remaining hot gas \cite{Bryan99,Voit00}. Finally, we have 
the ``star formation feedback'' sample, which is similar to the previous 
sample, except that newly formed stars return a fraction of their rest mass 
energy as thermal and mechanical energy. The source of this energy is high 
velocity winds and supernova energy from massive stars. In \textit{Enzo}, we implement 
this as thermal heating in every cell forming stars: $\Gamma _{sf} 
=\varepsilon _{SN} \dot {\rho }_{SF} c^2$. The feedback parameter depends on 
the assumed stellar IMF the explosion energy of individual supernovae. It is 
estimated to be in the range $10^{-6}\le \varepsilon _{SN} \le 10^{-5}$ \cite{Cen92}. 
We treat it as a free parameter.

Fig. \ref{fig10} shows synthetic maps of X-ray surface brightness, temperature, and 
Compton y-parameter for a $M=2\times 10^{15} M_{\odot}$ cluster at z=0 for the 
three cases indicated. The ``star formation'' case is omitted because the 
images are very similar to the ``star formation feedback'' case (see reference
\cite{Motl05}.) The adiabatic cluster shows that the X-ray emission is highly 
concentrated to the cluster core. The projected temperature distribution 
shows a lot of substructure, which is true for the adiabatic sample as a 
whole \cite{Loken02}. A complex virialization shock is toward the edge 
of the frame. The y-parameter is smooth, relatively symmetric, and centrally 
concentrated. The inclusion of radiative cooling has a strong effect on the 
temperature and X-ray maps, but relatively little effect on the SZE map. The 
significance of this is discussed in Section 5. In simulations with 
radiative cooling only, dense gas in merging subclusters cools to 10$^{4}$ K and 
is brought into the cluster core intact \cite{Motl04}. These cold lumps 
are visible as dark spots in the temperature map. They appear as X-ray 
bright features. The inclusion of star formation and energy feedback erases 
these cold lumps, producing maps in all three quantities that resemble 
slightly smoothed versions of the adiabatic maps. However, an analysis of 
the radial temperature profiles (Fig. \ref{fig10}) reveal important differences in 
the cluster core. The temperature continues to rise toward smaller radii in 
the adiabatic case, while it plummets to $\sim $10$^{4}$ K for the radiative 
cooling case. While the temperature profile looks qualitatively similar to 
observations of so-called cooling flow clusters, our central temperature is 
too low and the X-ray brightness too high. The star 
formation feedback case converts the cool gas into stars, and yields a 
temperature profile which follows the UTP at $r\ge 0.15r_{vir} $, but 
flattens out at smaller radii. This is consistent with the high resolution 
\textit{Chandra} observations of Vikhlinin et al. \cite{Vikhlinin04}.

\section{Recent Progress in Galaxy Cluster Modeling}

\begin{figure}
\begin{center}
\includegraphics[width=4.0in,height=3.5in]{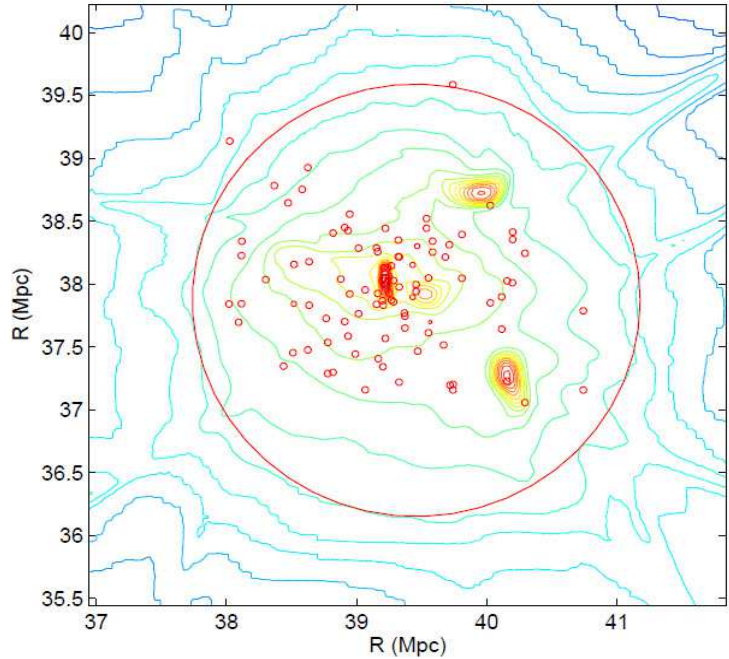}
\end{center}
\caption{Galaxy cluster simulation using {\em Galcons}. Galcons, short for
{galaxy constructs}, are programmable subgrid models for galaxies that have 
finite mass and size, are evolved kinematically as N-bodies, and feed back
mass, energy, metals, etc. to the ICM through a resolved spherical boundary at their
extremities. Small circles denote the location of 89 galcons, superimposed 
on a contour map of the baryon density. Large circle denotes the virial 
radius of the cluster. From \cite{Arieli08}.}
\label{fig.galcons}
\end{figure}

\begin{figure}
\begin{center}
\includegraphics[width=2.5in]{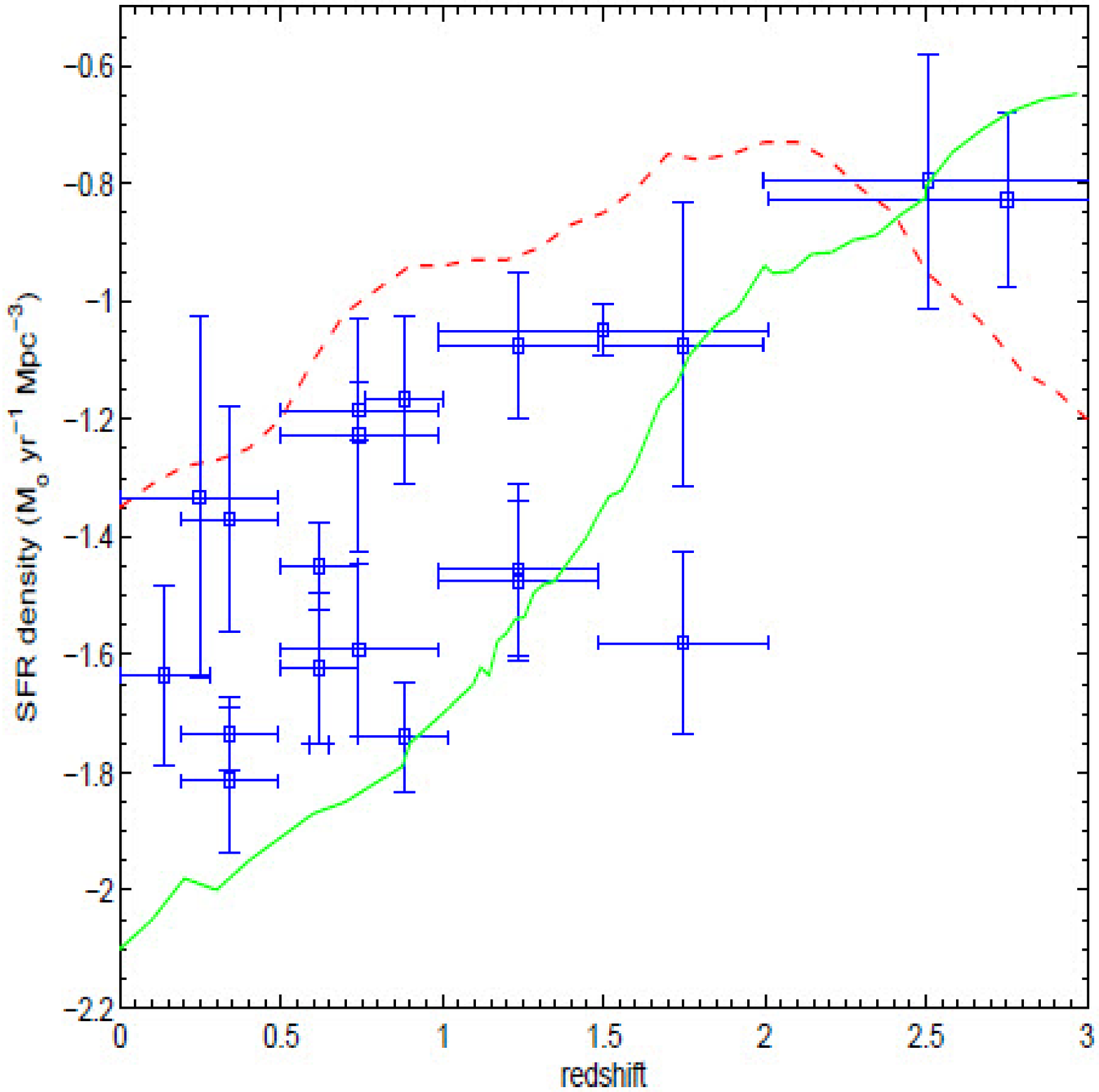}
\includegraphics[width=2.5in]{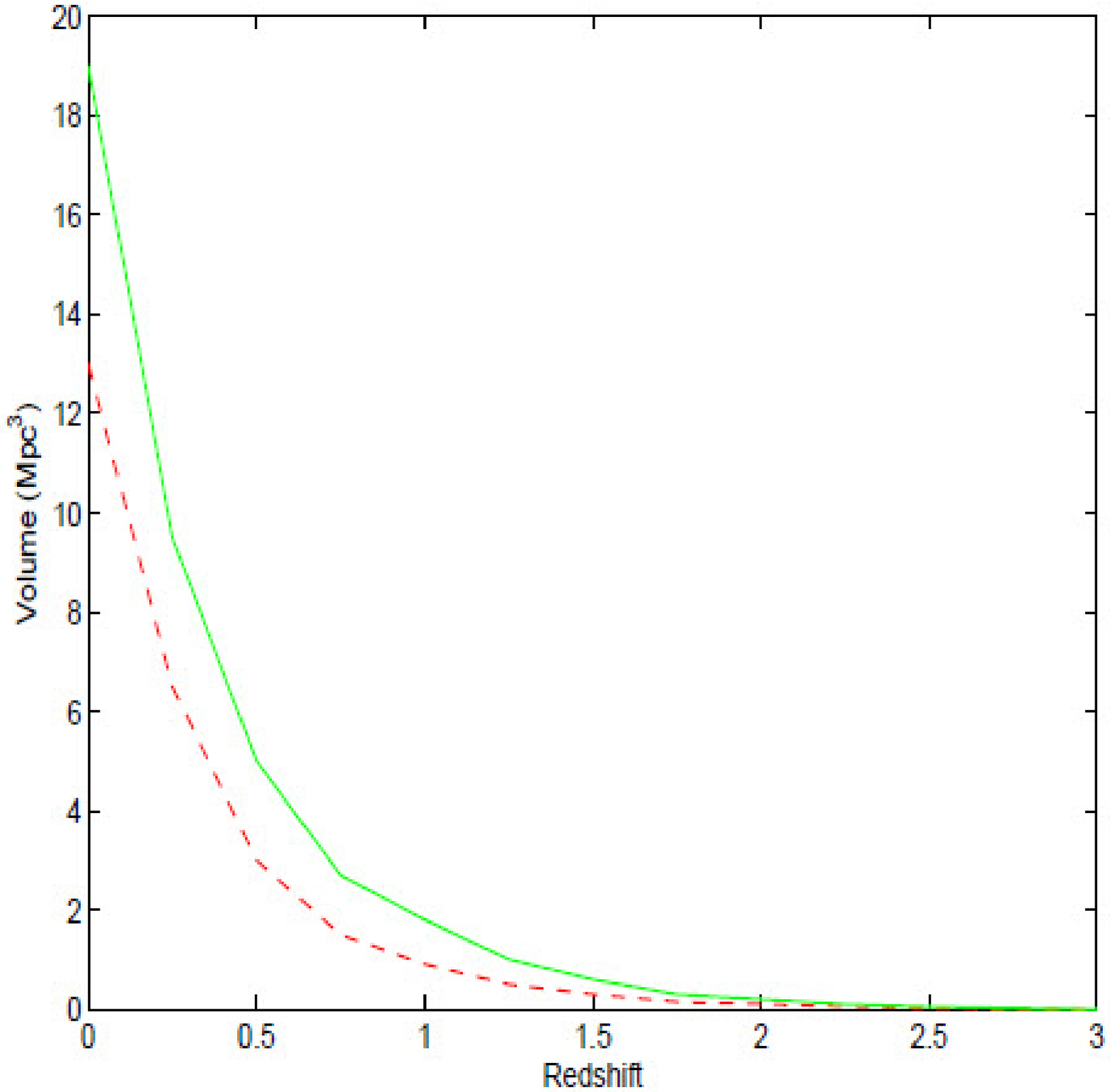}
\includegraphics[width=2.5in]{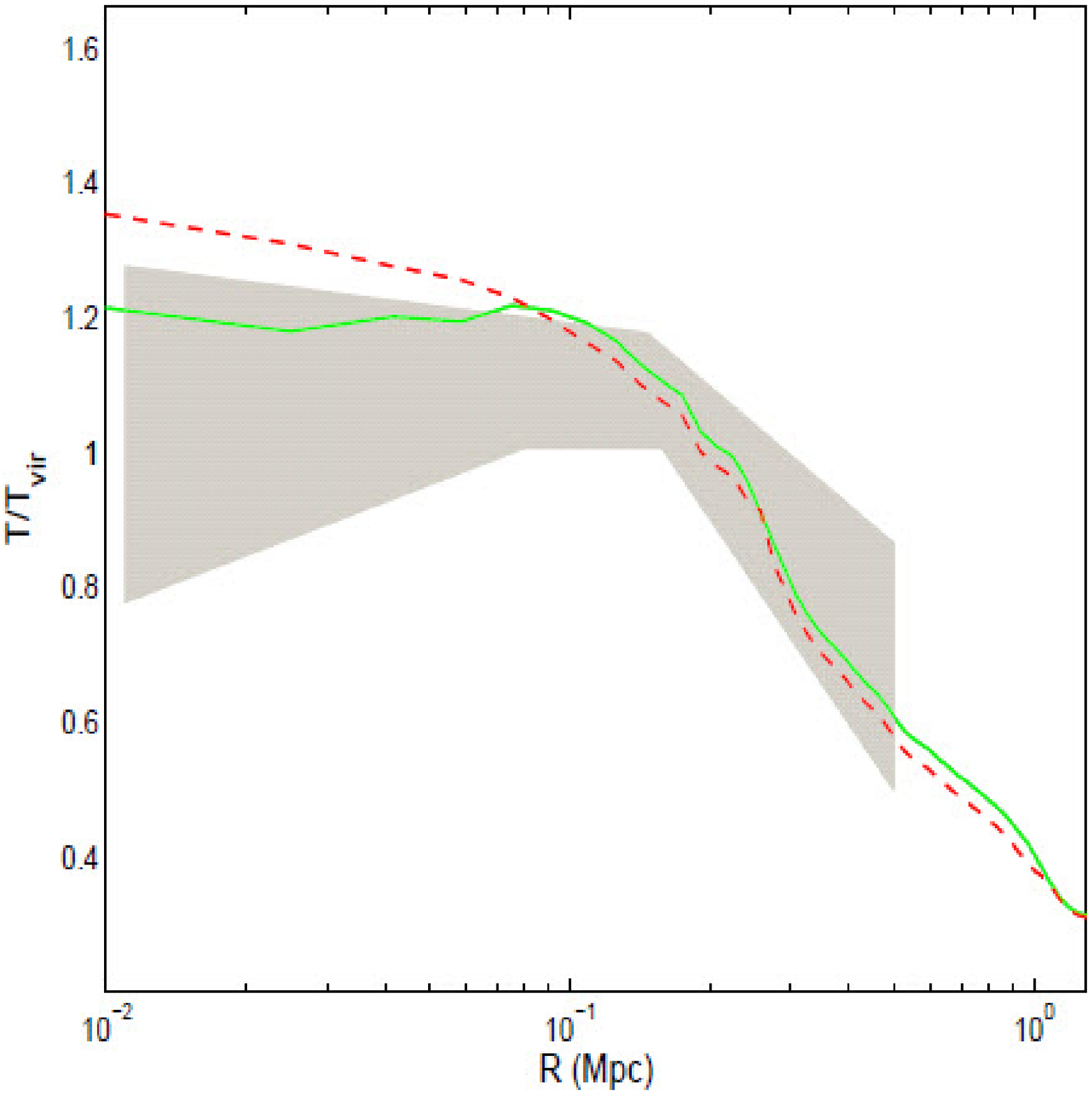}
\includegraphics[width=2.5in]{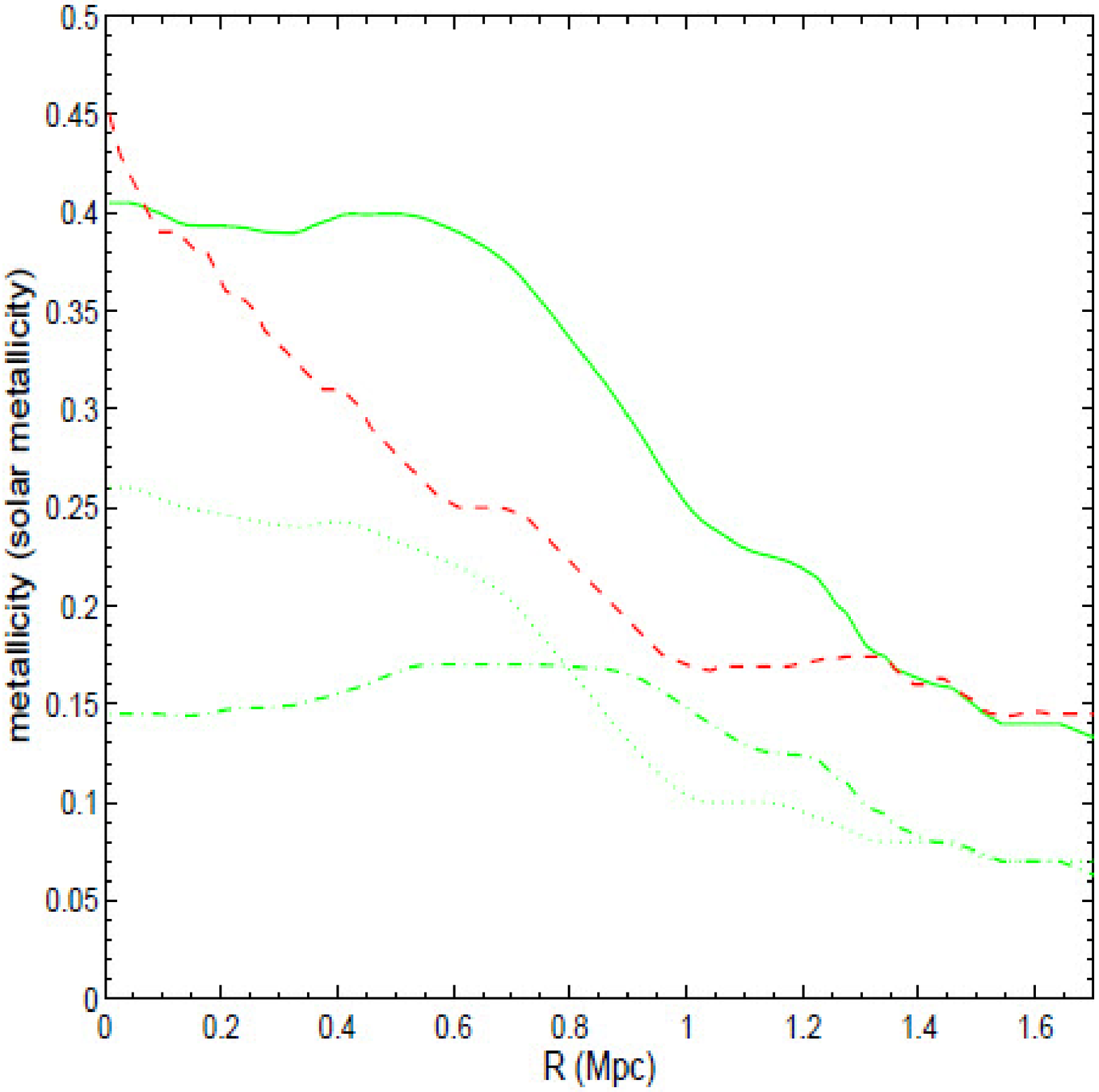}
\end{center}
\caption{Contrasting a {\em Galcon} cluster simulation (green solid lines)
with a simulation using
a well-known recipe for {\em in situ} star formation and feedback (red dashed lines).
Top left: star formation rate density versus redshift. Top right: volume of feedback
regions versus redshift. Bottom left: spherically averaged radial temperature profile. Shaded grey region bounds observed profiles from a sample of 15 clusters from \cite{Pratt07}. Bottom right: Metallicity profiles at $z=0$. Contributions to the total
metallicity from galactic winds (dashed-dotted green line) and due to ram pressure
stripping (dashed green line) are shown as well. From \cite{Arieli10}.}
\label{fig.galconresults}
\end{figure}

\subsection{Improved Treatment of Galaxy Feedback: GALCONS}
The problem with existing brute force approaches to modeling star formation 
and feedback within the context of galaxy cluster simulations is that they produce 
too few galaxies per cluster (unless extraordinarily high resolution is used), 
the galaxies form stars at too high a rate at late time 
(i.e., they are blue when they should be red), and the fraction of a cluster's baryonic mass 
in stars is too high \cite{Nagai07}. 
In this section we highlight a new approach to incorporating galaxies and their feedbacks
into hydrodynamic cosmological simulations of galaxy clusters which improves the agreement
between simulated and observed clusters \cite{Arieli08,Arieli10}. 
The key idea is rather than attempt to simulate the
internal processes of galaxies, these are modeled analytically or taken from observations. 
This is done through the introduction of a galaxy construct ({\em Galcon})--one per galaxy
dark matter halo--whose motion through the cluster is simulated as an N-body of finite
size and mass, but whose internal processes are programmable. 
The star formation history of each Galcon is an input function
rather than an output. This ensures agreement with observations. Feedbacks of energy and metals
is taken to be proportional the the instantaneous star formation rate, again calibrated by
observations. Feedback is done in a well-resolved spherical shell at the Galcon's outer radius rather
than from an unresolved point source, which ensures that the energy and metals get out into 
the IGM. Finally, and perhaps most importantly, a cluster with hundreds if not thousands
of galaxies can be simulated economically because the mass and spatial resolution requirement is not
so high. One requires only enough resolution to be able to find the galaxy dark matter halos
at high redshift which fall into the cluster. 

Fig. \ref{fig.galcons} shows the results of a Galcon simulation described in more detail in \cite{Arieli08,Arieli10}.
The procedure is as follows. We set up a standard Enzo simulation of a single galaxy cluster including
dark matter, gas but no radiative cooling and {\em in situ} star formation and feedback. In order to
achieve the required mass and spatial resolution in the cluster forming region we employ the nested grid+AMR 
strategy described in Sec. 4. We run this simulation to the ``replacement redshift" of z=3, where 
the output is analyzed for galaxy dark matter halos. For each halo we introduce a Galcon with an 
analytic mass model for stellar and gas distibution derived from a fit to the simulated baryon distribution.
We then restart the simulation and run it to z=0. We assume each Galcon's star formation history follows
the observationally determined cosmic star formation history with an appropriate weighting reflecting its mass. 
Mass loss through galactic winds and ram pressure stripping are modeled analytically. 

The upper left panel in Fig. \ref{fig.galconresults} compares the SFR density in the Galcon simulation with another Enzo simulation with identical
initial conditions but using the star formation and feedback recipe of Cen \& Ostriker (1992). 
Observational data points are also superposed. We see a sharply declining SFR in the Galcon simulation
after z=3, while the standard simulation rises and remains quite high to z=0 in conflict with observations.
Since feedback by galactic winds is assumed to be proportional to the SFR, this implies that the ICM in
the two simulations have very different heating histories. In the standard simulation, the heating is
confined to a small number of massive, centrally located galaxies late in time. In the Galcon simulation
heating occurs early in an extended region of space by nearly 100 galaxies before the cluster
collapses. This difference can be seen by comparing the redshift evolution of the 
volume of gas being heated by galaxy feedback (Fig. \ref{fig.galconresults}, upper right panel). 

The different heating histories and spatial distribution has an effect on the spherically averaged
gas temperature profiles at z=0, as shown in the lower left panel of Fig. \ref{fig.galconresults}. The Galcon simulation produces an isothermal core 
out to r=0.1r$_{vir}$, while the standard simulation produces a temperature profile which continues
to rise to smaller radii. The former is in better agreement with observations (shaded region), although still
a bit high in the core region, while the latter appears to be inconsistent with observations. 

The different feedback histories in the two simulations are also reflected in the different 
metallicity profiles at z=0 shown in lower right panel of Fig. \ref{fig.galconresults}.
The Galcon simulation produces a flat metallicity profile at $Z=0.4Z_{\odot}$ out to a radius
of 600 kpc, in good agreement with observations, while the standard simulation shows a sharply
declining metallicity gradient. This difference can be understood by looking at the conribution
of metals due to galactic winds and ram pressure stripping in the Galcon simulation, shown as
two separate curves in Fig. \ref{fig.galconresults}. A metallicity floor of about $Z=0.15Z_{\odot}$ is contributed
by galactic winds driven by early star formation when the galaxies where spatially extended. 
Ram pressure stripping removes metal enriched ISM gas from the Galcons and deposits it
preferentially in the central regions of the cluster at late times where the ICM gas is
denser. 

This model can be significantly improved upon in several ways. First, Galcons could be introduced
dynamically rather than at one time, and starting earlier. This would require running a halo finder 
inline with the calculation, which we can now do with Enzo \cite{Skory10}. 
Second, we could import star formation histories more appropriate to cluster galaxies, rather than 
assume the globally average rate. 
If one believes that Lyman break galaxies are the precursors of cluster galaxies \cite{Nagamine04},
then their high rates of star formation and strong outflows would provide more heating and hence
higher entropies in the cluster core than we have simulated. This could help establish the
observed entropy floor that current simulations fail to produce. 
Finally, we could allow our Galcons to merge using simple kinematic rules, which we presently
do not do. We could introduce a burst of star formation in a way which is consistent with
observations, and possibly even fuel AGN activity which could provide extra heating. 

\begin{figure}
\begin{center}
\includegraphics[width=\textwidth]{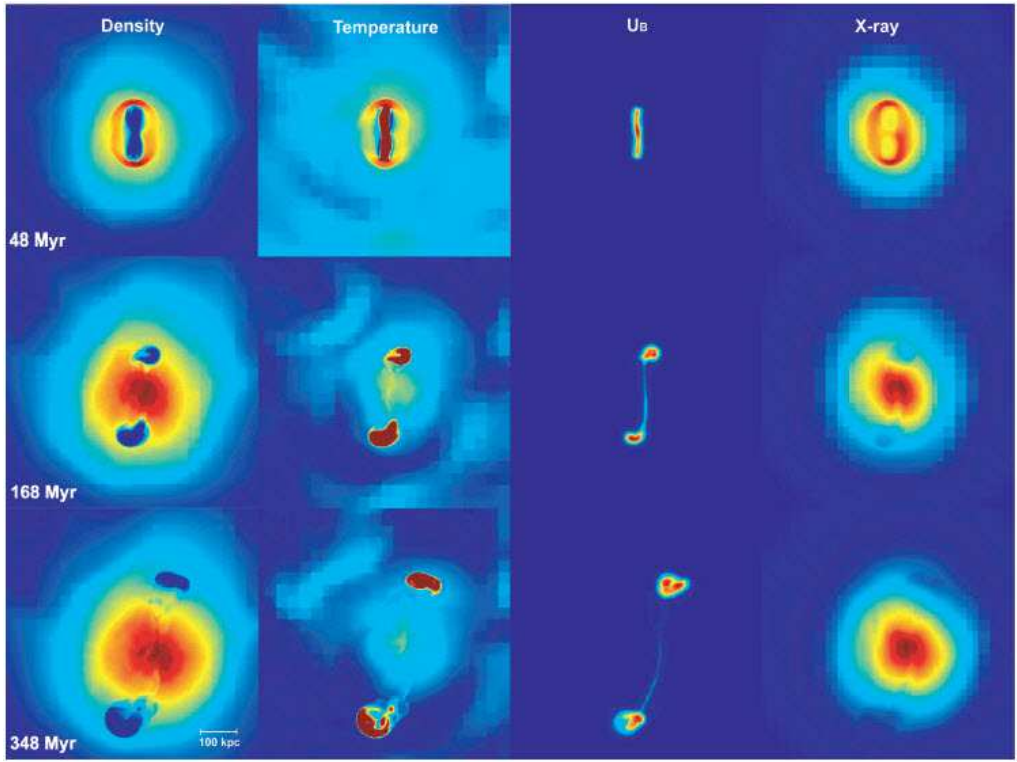}
\end{center}
\caption{Snapshots of the jet-lobe evolution in a realistic cluster ICM driven by the
magnetic energy output of an AGN. Each image is 672 kpc on a side. Columns from left
to right are slices through the cluster center of gas density, gas temperature, magnetic
energy density, and X-ray emissivity. Rows correspond to evolutionary times after the
turn-on of the AGN, which remains in the on state for 36 Myr. As can be seen, well defined
X-ray cavities with sharp, smooth boundaries are formed which remain intact for hundreds
of Myr as they bouyantly rise through the ICM. From \cite{Xu08}. }
\label{fig.jetlobe}
\end{figure}

\subsection{AGN Feedback: X-ray Cavity Formation by a Magnetized Jet}
High-resolution X-ray images of galaxy clusters by Chandra have revealed giant cavities and weak shock 
fronts in the hot gas 
\cite{Fabian00,McNamara00,McNamara05}
which are commonly associated with energetic radio lobes 
\cite{Blanton01,Nulsen02}
and suggest that magnetic fields play an important role. 
Numerical simulations of hot, underdense bubbles in
galaxy clusters have been performed by a number of authors 
\cite{Churazov01,Reynolds01,Bruggen02,Omma04}.
It is generally possible to inject a large amount of energy into 
the ICM via AGNs, but it is not exactly clear how the AGN energy can be efficiently utilized 
\cite{Vernaleo06}. 
One of the most interesting characteristics of the radio bubbles is that they 
are intact, whereas most hydrodynamic simulations 
\cite{Quilis01,Bruggen02,DallaVecchia04}
have shown that purely hydrodynamic bubbles will disintegrate
on timescales of much less than $10^8$ yr, markedly different from observations.

Using the first cosmological AMR MHD simulations, Xu et al. (2008)\cite{Xu08} have shown that intact X-ray cavities 
can be produced with properties similar to those observed \cite{Fabian00,McNamara00}. 
The simulations model the formation of a galaxy cluster within its proper cosmological context with magnetic
energy feedback from an active galactic nucleus (AGNs). The X-ray cavities are produced by
the magnetically dominated jet-lobe system that is supported by a central axial current. The cavities are magnetically
dominated, and their morphology is determined jointly by the magnetic fields and the background cluster
pressure profile. The expansion and motion of the cavities are driven initially by the Lorentz force of the magnetic
fields, and the cavities only become buoyant at late stages (1500 Myr). Interestingly, Xu et al.
find that up to 80--90\% of the injected magnetic energy goes into doing work against the hot cluster medium, 
heating it, and lifting it in the
cluster potential.

The simulation used the newly developed cosmological AMR-MHD module for the Enzo code described in
\cite{Collins10}. The procedure is evolve a galaxy cluster from cosmological initial conditions to z=0.05.
Magnetic feedback of a SMBH in the cluster center is modeled by injecting both poloidal and toroidal magnetic
flux in a divergence-free way into the central few kpc for $3 \times 10^7$ yr. The magnetic configuration is not 
force-free, but rather develops bipolar magnetically dominated ``towers" that extend along an axis. The
total amount of magnetic energy injected is $6 \times 10^{60}$ erg. After the source switches off, the cluster
is evolved for another 650 Myr.

Fig. \ref{fig.jetlobe} shows snapshots of the jet-lobe evolution driven by the magnetic energy output of an AGN. Each image is 
672 kpc  on the side. Columns from left to right are slices of density, temperature, the averaged magnetic 
energy density, and the integrated X-ray luminosity, respectively. The top row shows the cluster with
the jet lobe at the end of magnetic energy injection. The middle and bottom rows show the well-developed bubbles 
moving out of the cluster center. The bubbles are driven by magnetic forces at all stages and might become 
buoyant only after $t > 500$ Myr. As can be seen, the X-ray cavities remain intact with sharply defined boundaries
until the end of the simulation.

\begin{figure}[htbp]
\begin{center}
\includegraphics[width=0.9\textwidth]{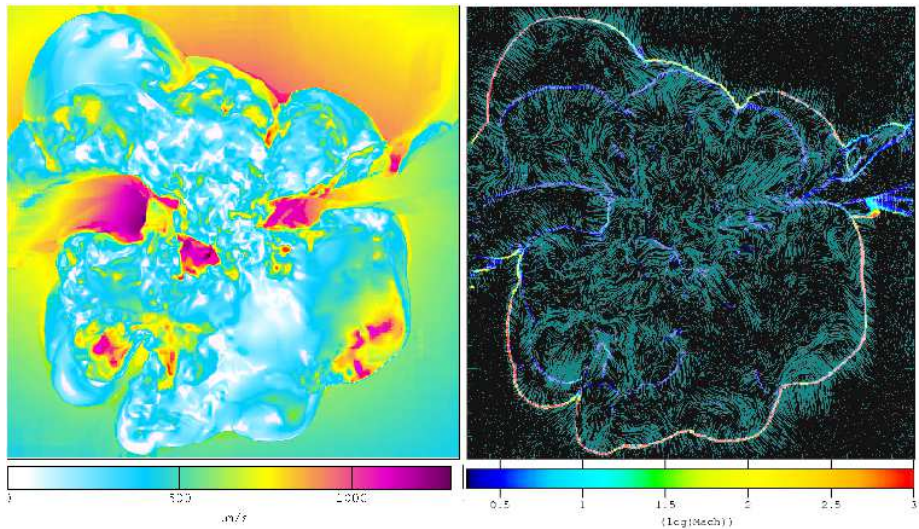}
\end{center}
\caption{Shocks and turbulence from an Enzo AMR simulation of a galaxy cluster
using a new AMR refinement criterion based on the velocity field. Left: magnitude
of the gas velocity on a slice 7.5 Mpc on a side with a resolution of 18 kpc
throughout the cluster. Right: map of Mach number (colors) and turbulent velocity
vectors on the slice. The virialization shock and internal shocks are kept
very sharp using the new refinement criterion. From \cite{Vazza09}.}
\label{fig.turbo_fig}
\end{figure}

\subsection{ICM Turbulence}
It is now appreciated that a non-negligible fraction of a cluster's binding energy is in the form of fluid
turbulence. Major mergers are the most likely source of energy, stirring the ICM on scales approching the 
virial radius. Direct evidence for cluster turbulence would be Doppler broading of X-ray lines,
or the kinetic SZ effect \cite{Sunyaev03}. As yet, these velocity signatures have not been detected
due to instrumental limitations. 
Indirect evidence is the strong magnetic fields observed in cluster cores via Faraday rotation measurements
(see discussion below). Here, strong fields would be the result of turbulent amplification of some small
seed field. 

Cluster turbulence was first studied by Norman \& Bryan (1999)\cite{BN99} using the newly developed ENZO AMR code. 
They simulated the formation of a rich cluster in a SCDM cosmological model assuming non-radiative
gas dynamics. They found turbulent velocities and bulk flows in the core of around 0.25 $\sigma_{vir}$,
increasing to as large as 0.6 $\sigma_{vir}$ near the virial radius. This translates into a turbulent
pressure of about 10\% of the total in the core region, increasing to about 30\% near the virial radius,
although distinguishing between what is an ordered versus a disordered flow becomes problematic near
the virial radius. These results were broadly confirmed through the Santa Barbara Cluster code 
comparison project, in which a number of different numerical methods were applied to evolve a common
set of initial conditions leading to the formation of a Coma-like cluster
\cite{Frenk99}.

\begin{figure}[htbp]
\begin{center}
\includegraphics[width=\textwidth]{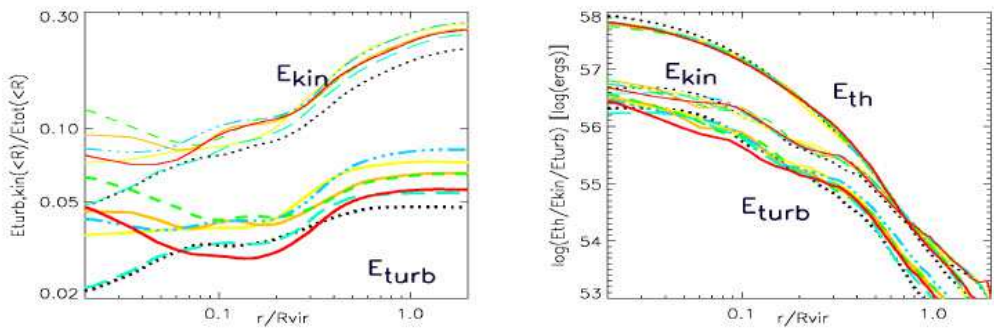}
\end{center}
\caption{Radial profiles of
turbulent kinetic, bulk kinetic, and thermal energies for the same cluster simulated with different resolutions and choices for the AMR refinement criterion. Left: as fractions of the total energy (turbulent+bulk+thermal). Right. Unnormalized profiles. From \cite{Vazza09}.}
\label{fig.turbo_profiles}
\end{figure}

Recently Vazza et al. (2009)\cite{Vazza09} have implemented a new adaptive mesh refinement criterion into Enzo based on
velocity jumps that provides higher resolution within the cluster virial radius and thus a better
characterization of the turbulent state of the ICM gas. They found an outer scale for the turbulence of about
300 kpc and a velocity power spectrum consistent with Kolmogorov for incompressible turbulence
$E_k \sim k^{-5/3}$. They found that compared to the standard density-based refinement criterion, 
their clusters have lower central gas densities, flatter entropy profiles, and a higher level 
of turbulence at all radii. The ratio of turbulent kinetic energy to thermal energy is
found to be $\sim 5$ percent within $0.1 R_{vir}$, increasing to  
$\sim 10-20$ percent within $R_{vir}$. The trend of increasing levels of turbulence with radius
is consistent with Norman \& Bryan (1999), however the absolute levels are somewhat lower
because Vazza et al. did a more careful job separating the velocity field into bulk and turbulent
components. They find bulk flows (velocity fields which are ordered on scales of 300 kpc or more)
also contribute substantially to the nonthermal energy budget, consistent with the findings of
Norman \& Bryan (1999).

Fig. \ref{fig.turbo_fig} shows the velocity field on a slice through the cluster center, and Fig. \ref{fig.turbo_profiles} shows radial profiles of
turbulent, bulk kinetic, and thermal energies for the same cluster simulated with different choices for the
AMR refinement criterion.

\begin{figure}
\begin{center}
\includegraphics[width=\textwidth]{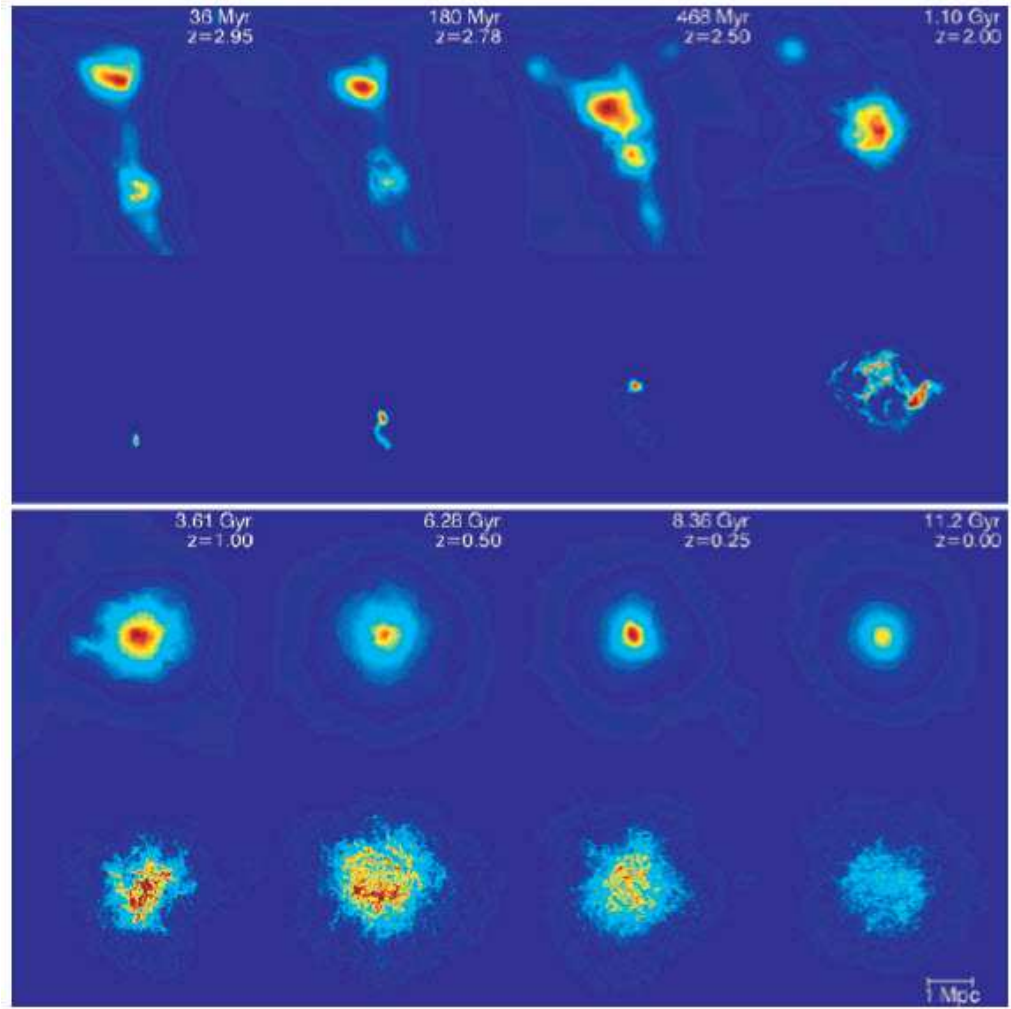}
\end{center}
\caption{Enzo AMR MHD cosmological simulation of the magnetization of the ICM
of a galaxy cluster due to a single AGN injection event at $z=3$. Snapshots of 
projected baryon density (upper rows) and magnetic energy density (bottom rows) for
different epochs of the evolving cluster. A major merger at $z=2$ induces cluster
turbulence which spreads and amplifies the magnetic field throughout the cluster
by $z=0$. This model predicts an RMS field strength which is nearly constant with
radius out to a Mpc at late times. From \cite{Xu09}. }
\label{fig.clusterB}
\end{figure}

\begin{figure}[htbp]
\begin{center}
\includegraphics[width=2.5in]{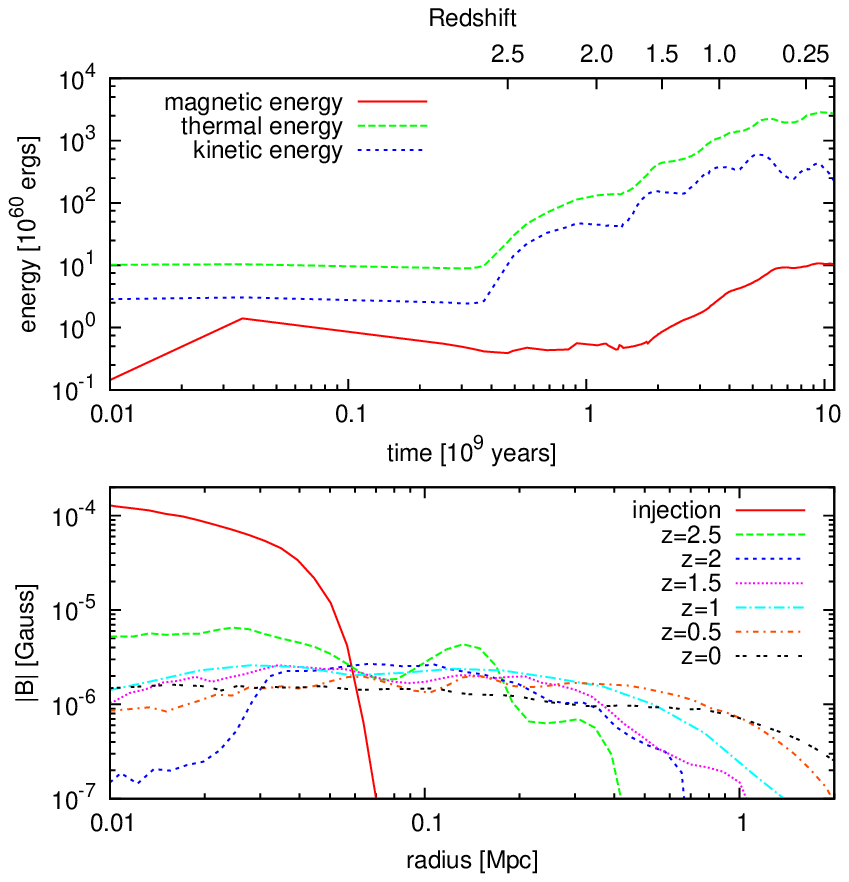}
\includegraphics[width=2.5in]{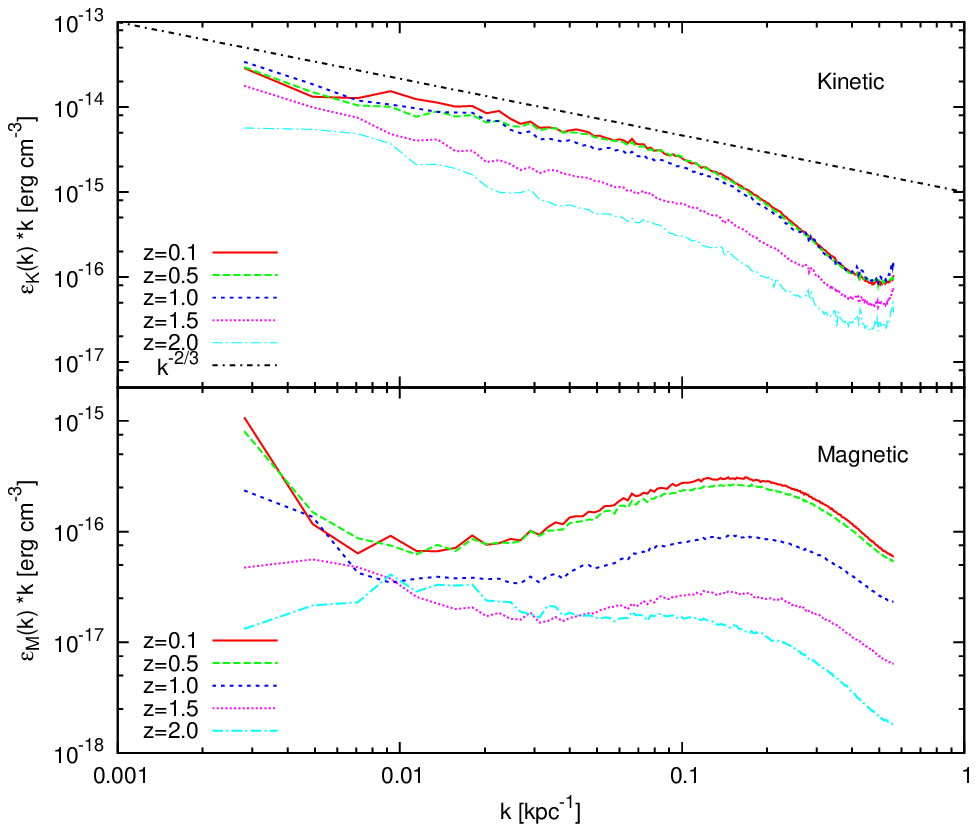}
\end{center}
\caption{Magnetic field evolution in the simulation shown in Fig. \ref{fig.clusterB}. Left: Magnetic energy versus redshift, as in integrated quantity (top) and as a function of radius. Right: Evolution of the power spectra of kinetic energy (top) and magnetic energy bottom).
From \cite{Xu09}}
\label{fig.Bevolution}
\end{figure}

\subsection{Origin of Cluster-Wide Magnetic Fields}
There is growing evidence that the intra-cluster medium (ICM)
is permeated with magnetic fields, as indicated by the detection
of large-scale, diffused radio emission called radio halos and
relics (see recent reviews by \cite{Ferrari08,Carilli02}. 
The radio emissions are extended over 1 Mpc,
covering the whole cluster. By assuming that the total energy in
relativistic electrons is comparable to the magnetic energy, one
often deduces that the magnetic fields in the cluster halos can
reach 0.1--1.0 $\mu$G and the total magnetic energy can be as high
as $10^{61}$ erg \cite{Feretti99}. The Faraday rotation measurement
(FRM), combined with the ICM density measurements, often
yields cluster magnetic fields of a few to 10 $\mu$G  (mostly in the
cluster core region). More interestingly, it reveals that magnetic
fields can have a Kolmogorov-like turbulent spectrum in the
cores of clusters \cite{Vogt03} with a peak at several
kpc. Other studies have suggested that the coherence scales of
magnetic fields can range from a few kpc to a few hundred
kpc \cite{Eilek02,Taylor93,Colgate01}, 
implying large amounts of magnetic energy and
fluxes. Understanding the origin and effects of magnetic fields
in clusters is important because they play a crucial role in
determining the structure of clusters through processes such
as heat transport, which consequently affect the applicability of
clusters as sensitive probes for cosmological parameters \cite{Voit05}.

The simulation described above suggests a way to magnetize the ICM of a galaxy cluster. 
Magnetic flux deposited into the IGM at high redshift by one or more AGN could provide a seed field
for subsequent turbulent amplification and mixing driven by cluster mergers. Norman \& Bryan (1999)
showed that cluster formation produces cluster-wide turbulence with Mach numbers ranging
from 0.1 in the core to 0.3 near the virial radius. This could drive a fast dynamo, amplifying
the seed field to observed levels. 

To test this hypothesis
Xu et al. (2009)\cite{Xu09} carried out a cosmological AMR MHD simulation using the ENZO+MHD code 
in which an AGN is switched on in a subcluster at a redshift of z=3. Just as in the
calculation described above, a magnetically-dominated jet-lobe system is formed by
injecting magnetic energy for 36 Myr. Thereafter the AGN is switched off, and the magnetic
fields evolve passively subject to the fluid dynamics of cluster assembly. 

Fig. \ref{fig.clusterB} shows images of the projected baryon density (upper row) and magnetic energy
density (bottom row) for various times in the evolution of the cluster. One can see that
magnetic energy, initially deposited in a small volume, is amplified and distributed 
throughout the cluster by the end of the simulation at z=0. 

Fig. \ref{fig.Bevolution} shows the evolution of the magnetic field in both time and space. A total of
about $2 \times 10^{60}$ erg of magnetic energy is injected over 36 Myr. It then 
decreases due to expansion to about 1/4 that value by z=1.5. Then it is amplified to
about $10^{61}$ erg by turbulence during cluster formation. The lower panel shows how
turbulent diffusion spreads the magnetic field throughout the cluster by z=0, with
an average value of about a microgauss out to a radius of 1 Mpc.

To see that turbulence is indeed the agent for field amplification and 
spreading, the right panel of Fig. \ref{fig.Bevolution} shows power spectra for gas kinetic energy and magnetic energy.
By forcing the AMR code to refine to the maximum level everywhere within the virial radius,
we have an effective uniform grid resolution of $600^3$ in the cluster region. 
We see a kinetic energy spectrum which is consistent with the Kolmogorov self-similar
scaling result over
the wavenumber range $0.01 \leq k \leq 0.1$ kpc$^{-1}$. Turbulent energy is
damped at smaller scales due to numerical dissipation, and back-reaction of
the magnetic field which is amplified the most at these scales.

\begin{figure}[htbp]
\includegraphics[width=\textwidth]{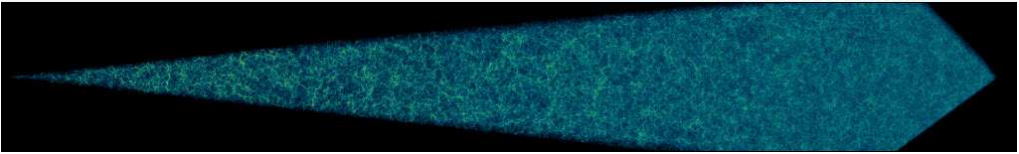}
\caption{Growth of cosmic structure ``on the lightcone", derived from the Hubble Volume Simulation \cite{HubbleVolume}.}
\label{fig.lightcone}
\end{figure}

\section{Statistical Ensembles and Lightcones}

With a volume as large as the Hubble Volume simulation, 
and an adequate number of intermediate snapshots in time,
it is possible to portray the growth of comic structure ``on the lightcone"; that is,
just as we observe the real universe. A graphical depiction of that is shown in Fig. \ref{fig.lightcone};
here only a thin slice through the narrow 3D lightcone is shown. Here one sees the time
evolution of the galaxy cluster population over the redshift interval $0 \leq z \leq 3$. 
Despite the low spatial and mass resolution of the simulation, it is plain that
cluster size objects appear rather late in the evolution of the universe.

\begin{figure}[htbp]
\includegraphics[width=2.5in]{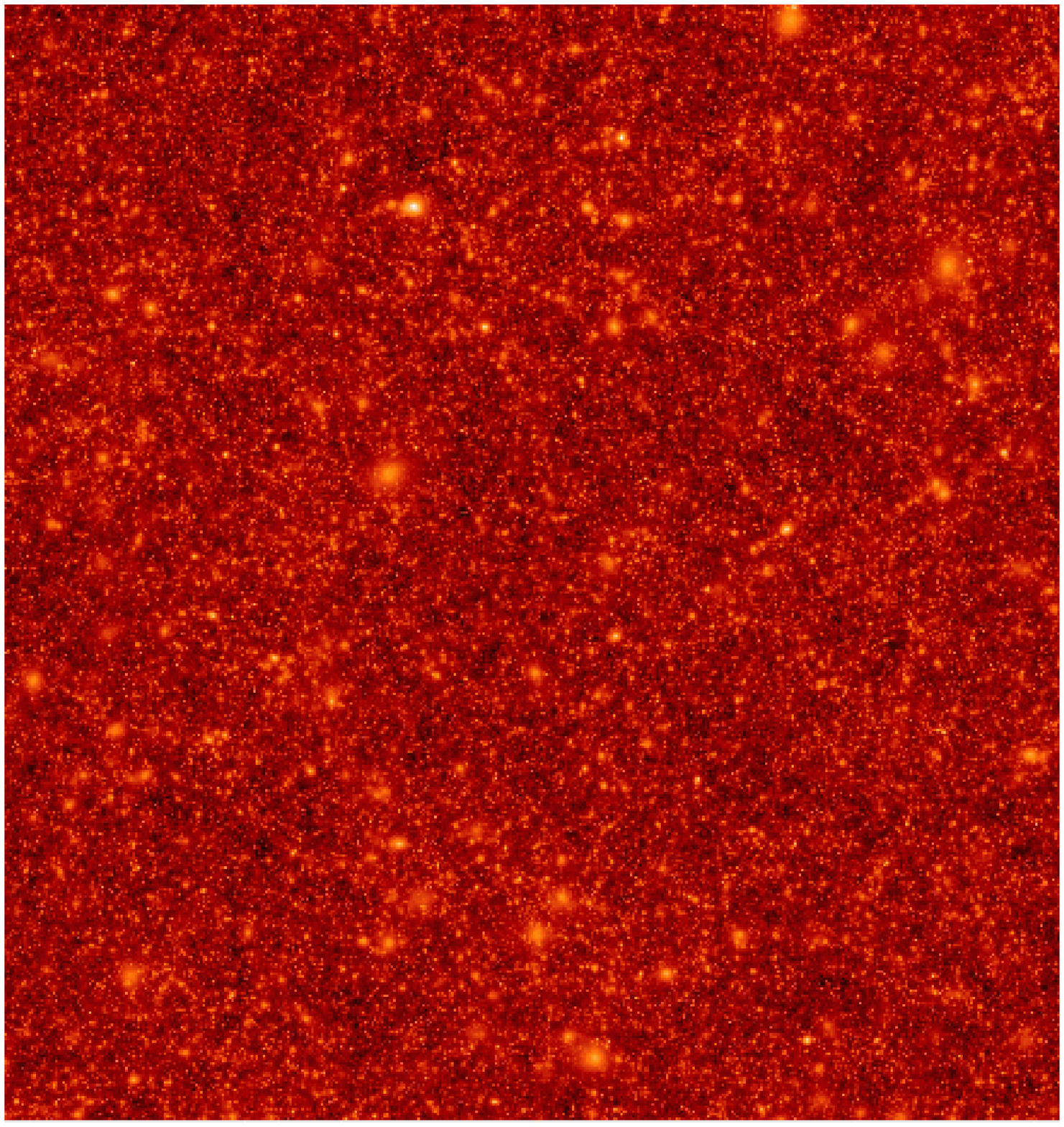}
\includegraphics[width=2.5in]{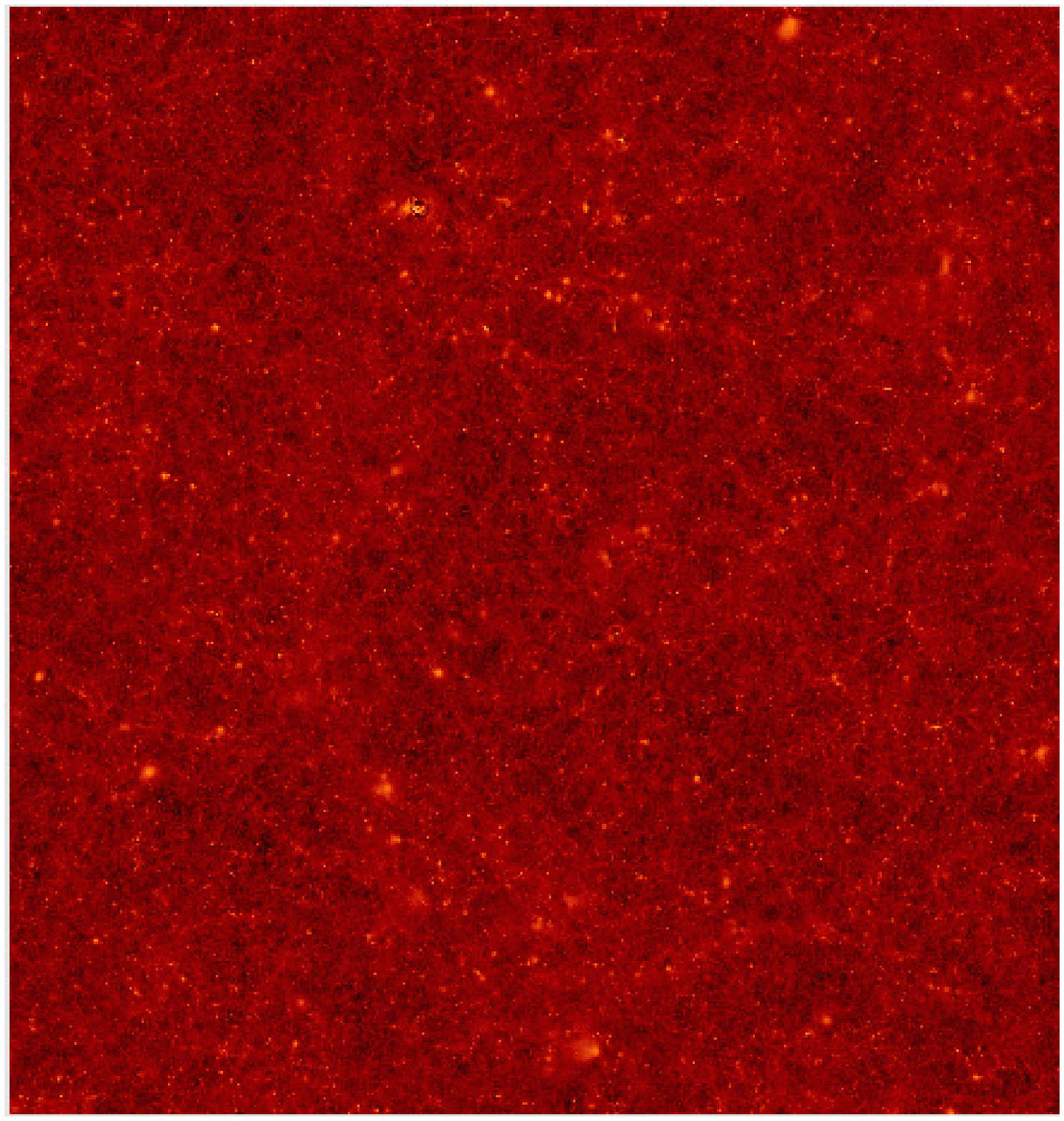}
\caption{Left: 100 deg$^2$ projection lightcone image of the Compton y-parameter from
a $512^3$ the Enzo AMR simulation shown in Fig. 1. 
Right: same as left panel, except removing the contribution of all virialized gas 
inside clusters of mass $M \geq 5 \times 10^{13} M_{\odot}$. From \cite{Hallman07}}
\label{fig.SZmap}
\end{figure}

At present, it is not possible to simulate volumes this large and at the same time
resolve the internal structure of galaxy clusters including the baryonic component. 
However, by simulating somewhat smaller (but still large) volumes, and exploiting
the periodicity of the boundary conditions, it is still possible to generate lightcones
for the purposes of synthetic deep redshift cluster surveys. An example of this is shown in
Fig. \ref{fig.SZmap} and described in more detail in Hallman et al. (2007)\cite{Hallman07}. It shows a 
100 square degree projected lightcone image of the Compton y parameter due to the
thermal SZE from galaxy clusters over the redshift interval $0.1 \leq z \leq 3$. 
In order to produce this image, 27 redshift outputs from the AMR hydrodynamic 
simulation shown in Fig. 1. are stacked at
$\Delta z$ intervals of 0.1. The simulation has a dark matter particle mass of  
$7.3 \times 10^{10} M_{\odot}$ and a maximum spatial resolution of $7.8 h^{-1}$
comoving kpc. This allows a fairly complete sample of clusters above $M_{halo}
\approx 4 \times 10^{13} M_{\odot}$ (Fig. \ref{fig.massfn}). 
In order to avoid repeating structures, each redshift ``shell" of the lightcone is
generated by projecting through the cube along a different random axis, and then
exploiting periodicity to shift and fill the required solid angle. Using different
random seeds for orientation and shift, multiple lightcone realizations can be
made from a single simulation in order quantify sampling errors due to cosmic
variance.  

\begin{figure}[htbp]
\includegraphics[width=\textwidth]{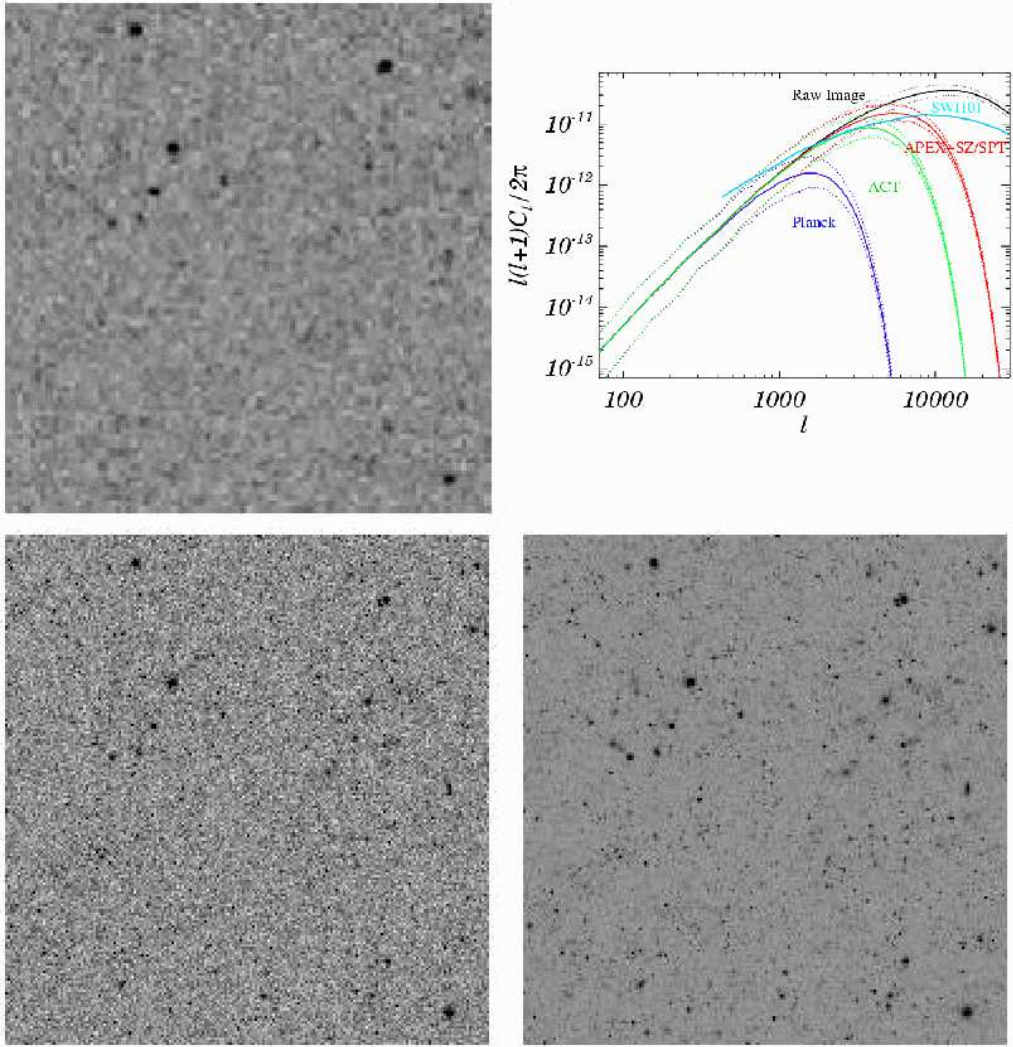}
\caption{Top Left: 100 deg$^2$ projection lightcone image of the Compton y-parameter modified
to model angular resolution and sensitivity of the {\em Planck Surveyer} all sky survey 
at 144 GHz. Top Right: Angular power spectrum generated from these images. 
Bottom Left: same as image in top left, except for Apex-SZ and SPT survey characteristics.
Bottom Right: same as image in top left, except for ACT survey characteristics.
From \cite{Hallman07}}
\label{fig.SZobs}
\end{figure}

\subsection{Mock SZE Surveys}
With such lightcone datasets, it is possible to mock up the blind SZE surveys 
that are being carried out by Planck, ACT, and SPT/APEX-SZ \cite{Hallman07}. First, an
observing frequency is chosen. Then an image similar to that shown in Fig. \ref{fig.SZmap}
is generated using the approprate frequency-dependent SZE signal (see Rephaeli, these
proceedings). Second, the image is convolved with the PSF of the relevant instrument. 
For the instruments listed above, the beam size assumed is 7.1, 1.7, and 1 arcmin, 
respectively at 144 GHz. Finally, noise is added to the image at the level of the
design sensitivities for the instruments, which are assumed to be 6, 2 and 10
$\mu K$ per beam, respectively. These images are shown in Fig. \ref{fig.SZobs}. The graph
at the upper right hand corner of Fig. \ref{fig.SZobs} shows the angular power spectrum 
computed for the three instruments, as well as the raw image before degrading 
resolution and sensitivity. The solid lines show the average power spectrum over
200 lightcone realizations, while the dotted lines show the range in which
90\% of the power is found.  One can see the effects of finite angular resolution
by the sharp turnover of the angular power at high multipoles. All instruments
recover the ``theoretical" power at low multipoles, however. Cosmic variance 
can change the measured power by factors of 5-8, implying that converged
measurements of the real power will require very large areas of the sky to be 
mapped. 

\begin{figure}[htbp]
\includegraphics[width=2.5in]{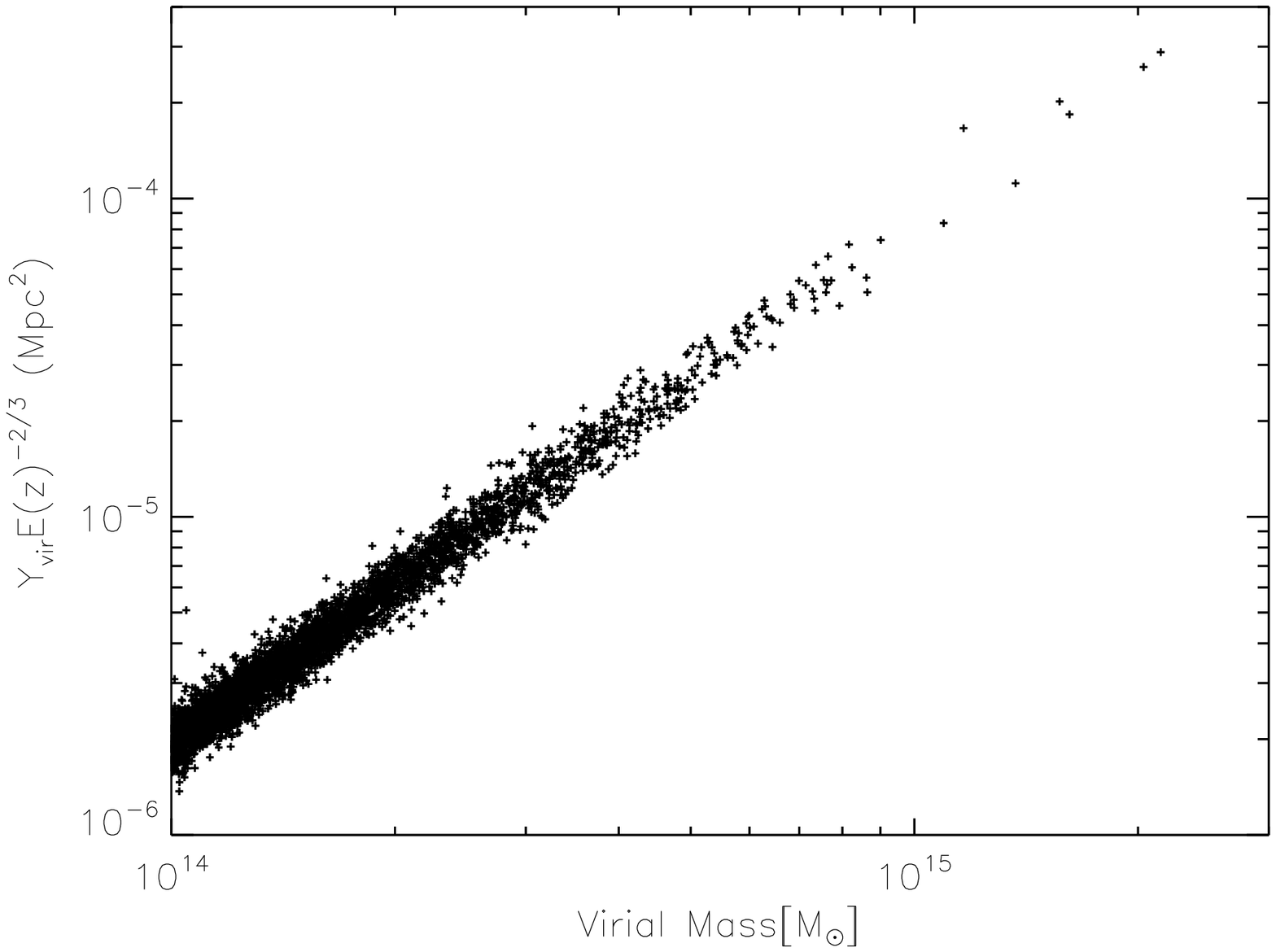}
\includegraphics[width=2.5in]{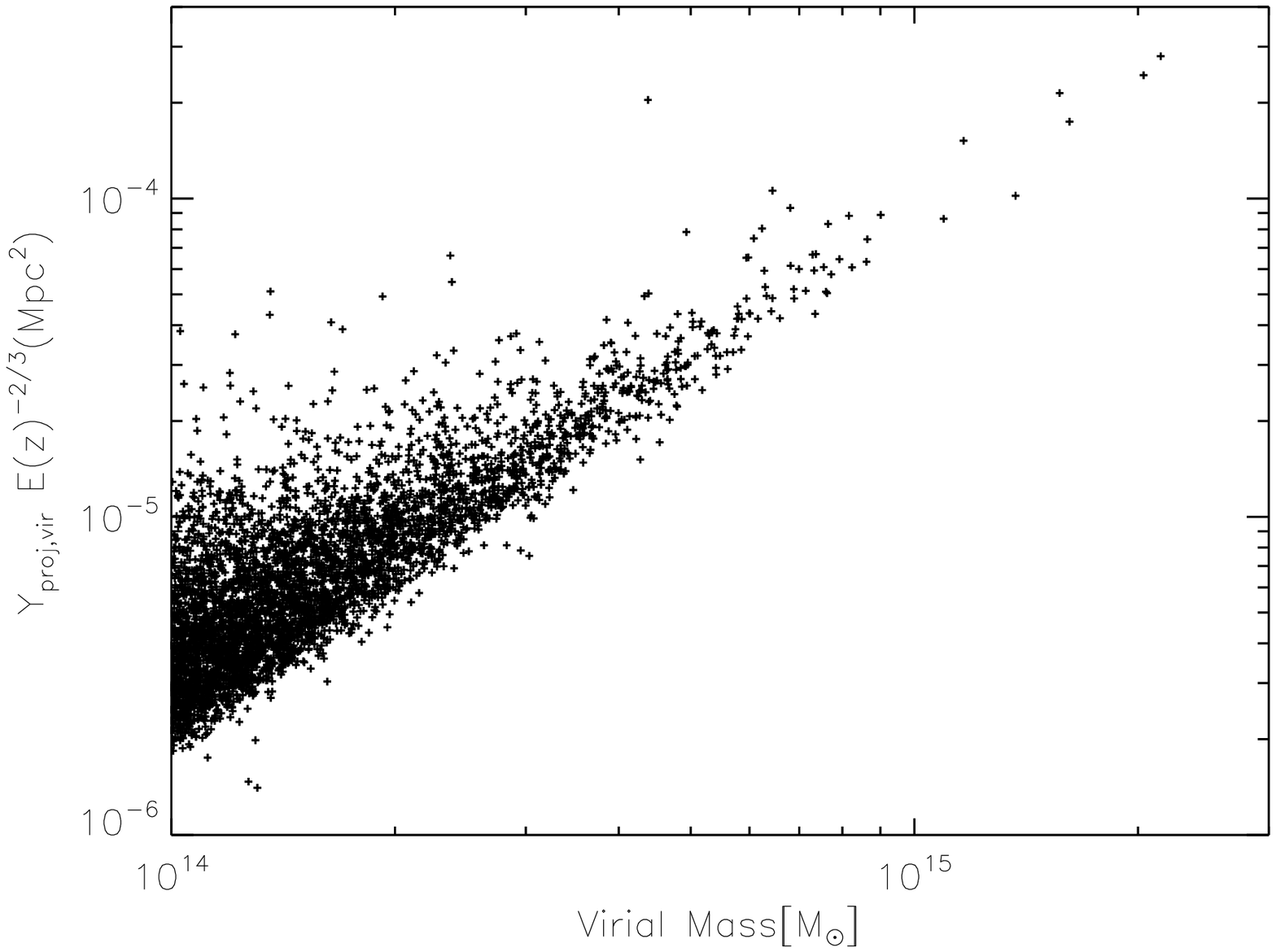}
\caption{Left: Scatter plot of Y-M relation (corrected for cosmological evolution)
taken directly from simulation data. Right: same as left panel except taken from
lightcone projections which include SZE signal from hot gas along the LOS to the
cluster. Tight correlation is severely degraded by projection effects. 
From \cite{Hallman07}}
\label{fig.scatterY}
\end{figure}

With simulated lightcones, one can ask whether projection effects damage the potential
of SZE surveys to constrain cosmological parameters. As discussed in Sec. 3.6
above, counting clusters down to a limiting mass/flux versus redshift is a
strong probe of cosmology, especially the parameters $\Omega_m$ and $\sigma_8$.
This requires calibrating the Y-M relation, where Y is the integrated SZE
and M is the cluster mass. Motl et al. (2005)\cite{Motl05} and Nagai (2006)\cite{Nagai06} showed using
hydrodynamic cosmological simulations that there is a tight correlation between
these two quantities that is rather insensitive to assumed baryonic processes
(cooling, star formation, feedback, etc.) However, because the SZE signal is
redshift independent, when galaxy clusters are
observed projected against the sky, foreground objects such as other clusters,
hot gas outside the virial radius, and Warm-Hot Intergalactic Medium (WHIM)
may contribute to the total signal and spoil the tight correlation. Hallman et al.
(2007) showed that indeed this is the case, as indicated in Fig. \ref{fig.scatterY}. 
On the left
is the Y-M relation scatter plot deduced from the simulation data itself, while
on the right we see the clusters in projection. The tight correlation intrinsic to
individual clusters is destroyed by other contributions to the SZ signal along
the LOS, especially for the more numerous low mass clusters.  As visual proof
of this we show in the right panel of Fig. \ref{fig.SZmap}
the projected SZ signal removing the contribution
of all cluster gas inside the virial radii of every cluster with
$M \geq 5 \times 10^{13} M_{\odot}$. Roughly 1/3 of the flux of the original
image remains after this subtraction is done. A majority of it arises in the
hot gas immediately outside the virial radius of clusters which was heated by
prior epochs of structure formation, while some comes from the WHIM \cite{Hallman09}. This signal is entirely absent
in analytic or semi-analytic models of the SZE in which the baryons are painted on 
a halo population taken from a Press-Schechter analysis or pure dark matter
N-body simulation. This highlights the importance of performing 
self-consistent hydrodynamic simulations and including the effects of foregrounds
through detailed lightcone analyses.

\acknowledgments
The author is indebted to his collaborators Greg Bryan, Jack Burns, David Collins, 
Claudio Gheller, Eric Hallman, Hui Li, Chris Loken, Patrick Motl, Franco Vazza, and
Hao Xu whose results, both 
published and unpublished, is presented here. Simulations were performed at 
the San Diego Supercomputer Center at the University of California, San Diego
with partial support of grants NSF AST-0708960 and AST-0808184.

\end{document}